\documentstyle[preprint,aps,epsf]{revtex}

\tightenlines

\begin{document}

\newcommand{\notE}{\ \hbox{{$E$}\kern-.60em\hbox{/}}}
\newcommand{\notp}{\ \hbox{{$p$}\kern-.43em\hbox{/}}}
\def\D0{\mbox{D\O}}


\preprint{\tighten
\font\fortssbx=cmssbx10 scaled \magstep2
\hbox to \hsize{
\hbox{\fortssbx University of Wisconsin - Madison}
\hfill$\vcenter{\hbox{\bf Fermilab-Pub-98/342-T}
                \hbox{\bf MADPH-98-1085}
                \hbox{\bf hep-ph/9811489}}$ }
%
}

\title{\vspace*{0.7cm}
Trilepton Signature of Minimal Supergravity \\
at the Upgraded Tevatron}

\author{V. Barger$^{1,2}$ and Chung Kao$^2$}

\address{
$^1$Fermi National Accelerator Laboratory,
P.O. Box 500, Batavia, IL 60510 \\
$^2$Department of Physics, University of Wisconsin,
Madison, WI 53706}

\maketitle

\thispagestyle{empty}

\begin{abstract}

The prospects for detecting trilepton events ($\ell = e$ or $\mu$)
from chargino-neutralino ($\chi^\pm_1 \chi^0_2$) associated production
are investigated for the upgraded Fermilab Tevatron Collider
in the context of the minimal supergravity model (mSUGRA).
In some regions of parameter space, $\chi^\pm_1$ and $\chi^0_2$
decay dominantly into final states with $\tau$ leptons
and the contributions from $\tau-$leptonic decays enhance
the trilepton signal substantially when soft cuts
on lepton transverse momenta are used.
Additional sources of the mSUGRA trilepton signal
and dominant irreducible backgrounds are discussed.
The dilepton ($\ell^+\ell^-$) invariant mass distribution near the endpoint
is considered as a test of mSUGRA mass relations.
Discovery contours for $p\bar{p} \to 3\ell +X$ at 2 TeV
with an integrated luminosity of 2 fb$^{-1}$ to 30 fb$^{-1}$
are presented in the mSUGRA parameter space of $(m_0,m_{1/2})$
for several choices of $\tan\beta$.

\end{abstract}

%


\section{Introduction}


In the near future the Main Injector (MI) of the Fermilab Tevatron Collider
will run at 2 TeV center of mass energy with a luminosity
of about $10^{32}$ cm$^{-2}$ s$^{-1}$ and will accumulate
an integrated luminosity (${\cal L}$) of 2 fb$^{-1}$ (Run II)
or more at each of the CDF and the \D0 detectors.
It has been proposed to further upgrade the Tevatron luminosity to
$10^{33}$ cm$^{-2}$ s$^{-1}$ to obtain a combined integrated luminosity
${\cal L} = 30$ fb$^{-1}$ (Run III) \cite{Gilman,TeV2000}.
Another possibility is that the MI will run at the Run II luminosity
for more years to accumulate a higher integrated luminosity.
The order of magnitude increase in luminosity beyond the 0.1 fb$^{-1}$
\cite{D0,CDF} now available will significantly improve the possibility
that new physics beyond the Standard Model (SM) could be discovered
before the CERN Large Hadron Collider (LHC) begins operation \cite{TeV2000}.

In this article we extend our recent study \cite{Madison}
on the prospects of detecting the trilepton signal
with missing transverse energy
in the minimal supergravity unified model (mSUGRA)
at the upgraded Tevatron with a detailed consideration
of backgrounds and optimized acceptance cuts
to improve the trilepton search.
The primary source of trileptons is associated production of
the lighter chargino ($\chi^\pm_1$)
and the second lightest neutralino ($\chi^0_2$) with both decaying
to leptons \cite{Trilepton1,Trilepton2,Trilepton3,BCDPT2}.
In mSUGRA with gauge coupling unification,
the sleptons ($\tilde{\ell}$), the lighter chargino ($\chi^\pm_1$)
and the lighter neutralinos ($\chi^0_1,\chi^0_2$)
are considerably less massive than the gluinos and squarks
over most of the parameter space.
Because of this, the trilepton signal ($3\ell+\notE_T$)
is the most promising channel
\cite{D0,CDF,Madison,Trilepton1,Trilepton2,Trilepton3,BCDPT2}
for supersymmetric particle searches at the Tevatron.
The trilepton background from SM processes can be greatly reduced
with suitable cuts.

In supergravity unified models \cite{SUGRA},
supersymmetry (SUSY) is broken in a hidden sector
with SUSY breaking communicated to the observable sector
through gravitational interactions,
leading naturally but not necessarily \cite{Non-universal}
to a common scalar mass ($m_0$), a common gaugino mass ($m_{1/2}$),
a common trilinear coupling ($A_0$) and a bilinear coupling ($B_0$)
at the grand unified scale ($M_{\rm GUT}$).
Through minimization of the Higgs potential,
the $B$ coupling parameter of the superpotential
and the magnitude of the Higgs mixing parameter $\mu$ are related to
the ratio of Higgs-field vacuum expectation values (VEVs)
($\tan\beta \equiv v_2/v_1$) and to the mass of the $Z$ boson ($M_Z$).
The SUSY particle masses and couplings at the weak scale can be predicted
by the evolution of renormalization group equations (RGEs) \cite{RGE}
from the unification scale \cite{BBO,SUGRA2}.
We evaluate SUSY mass spectra and couplings
in the minimal supergravity model
in terms of $m_0$, $m_{1/2}$, $A_0$ and $\tan\beta$,
along with the sign of the Higgs mixing parameter $\mu$.
The value of $A_0$ does not significantly affect our analysis;
therefore, we take $A_0 = 0$ in our calculations.
Non-universal boundary conditions among sfermion masses \cite{sfermions}
or the gaugino masses \cite{gauginos}
could change the production cross section and branching fractions
of the charginos and neutralinos.
For $m_{1/2} =$ 200 GeV and $\tan\beta \alt 25$,
a non-universality among sfermions significantly enhances
the trilepton signal when 50 GeV $\alt m_0 \alt$ 130 GeV \cite{sfermions}.

The mass matrix of the charginos in the basis of the weak eigenstates
($\tilde{W}^\pm$, $\tilde{H}^\pm$) has the following form
\begin{equation}
M_C=\left( \begin{array}{c@{\quad}c}
M_2 & \sqrt{2}M_W\sin \beta \\
\sqrt{2}M_W\cos \beta & \mu
\end{array} \right)\;.
\label{eq:xino}
\end{equation}
Since this mass matrix is not symmetric, its diagonalization requires
two matrices \cite{MSSM}.
The sign of the $\mu$ contribution in Eq.~(\ref{eq:xino})
establishes our sign convention\footnote{
This sign convention for $\mu$ is opposite to that of Ref.~\cite{Madison}.}
for $\mu$, which is equivalent to the ISAJET convention \cite{ISAJET}.

In Fig. 1, we present the masses of
the lightest neutralino ($\chi^0_1$),
the second lightest neutralino ($\chi^0_2$),
the lighter chargino $\chi^\pm_1$,
the scalar electrons $\tilde{e}_L$ and $\tilde{e}_R$,
the lighter tau slepton ($\tilde{\tau}_1$),
and the lighter bottom squark $\tilde{b}_1$
at the mass scale of $M_Z$,
and the mass of the lighter CP-even Higgs scalar ($h^0$) at the scale
$Q = \sqrt{  m_{\tilde{t}_L} m_{\tilde{t}_R} }$ \cite{BCDPT1,Relic},
versus $m_0$,
with $M_{\rm SUSY} =$ 1 TeV, $m_{1/2} = 200$ GeV and $\mu > 0$
for (a) $\tan\beta = 2$ and (b) $\tan\beta = 35$.
To a good approximation,
the mass of the lightest chargino $m_{\chi^\pm_1} \sim m_{\chi^0_2}$
is about twice $m_{\chi^0_1}$.
Also shown in Fig. 1 are the regions that do not satisfy
the following theoretical requirements:
tachyon free and the lightest neutralino
as the lightest SUSY particle (LSP).
There are several interesting aspects to note in Fig. 1: \\
(i) An increase in $\tan\beta$ leads to a larger $m_h$
but a slight reduction in $m_{\chi^0_1}$, $m_{\chi^\pm_1}$,
and a large reduction of $m_{\tilde{\tau}_1}$ and $m_{\tilde{b}_1}$. \\
(ii) Increasing $m_0$ raises the masses of scalar fermions. \\
(iii) In most of the mSUGRA parameter space,
the weak-scale gaugino masses are related
to the universal gaugino mass parameter $m_{1/2}$ by
\begin{eqnarray}
m_{\chi^0_1}   & \sim & 0.44 m_{1/2}, \;\; {\rm and} \nonumber \\
m_{\chi^\pm_1} & \sim & m_{\chi^0_2} \sim 0.84 m_{1/2}.
\end{eqnarray}
Consequently, the trilepton channel
could provide valuable information about the value of $m_{1/2}$.

The masses of $\chi^\pm_1$ and $\chi^0_2$ to the leading order
in $M_W^2/(\mu^2-M_2^2)$ can be expressed as \cite{Mass}
\begin{eqnarray}
m_{\chi^\pm_1} & = & M_2 -\frac{ M_W^2(M_2+\mu\sin 2\beta) }
                               { \mu^2-M_2^2+2M_W^2 }, \nonumber \\
m_{\chi^0_2}   & = & M_2 -\frac{ M_W^2(M_2+\mu\sin 2\beta) }
                               { \mu^2-M_2^2 }
\end{eqnarray}
We find that $M_2$ and $|\mu|$ can be empirically expressed in GeV units
as a function of the GUT scale masses $m_{1/2}$, $m_0$ and $\cos 2\beta$ as
\begin{eqnarray}
M_2   & = & 0.851 m_{1/2} +0.00244 m_0 -2.20, \nonumber \\
|\mu| & = & a m_{1/2} +b \cos 2\beta +c, \nonumber \\
    a & = &   2.34 -0.153 (m_0/100\;{\rm GeV})
           +[ 1.10 -0.141 (m_0/100\;{\rm GeV}) ] \cos 2\beta, \nonumber \\
    b & = & 1.787 m_0 -167.5, \nonumber \\
    c & = & 1.909 m_0 -178.7,
\end{eqnarray}
for 100 GeV $\alt m_0, m_{1/2} \alt$ 1000 GeV and all $\tan\beta$
for which perturbative RGE solutions exist.
These mass formulas hold
to an accuracy of $5\%$ for $M_2$, $m_{\chi^\pm_1}$ and $m_{\chi^0_2}$, and
to an accuracy of $10\%$ for $|\mu|$.
These formulas provide useful approximations for quick estimates.
In our analysis, we have used the numerical results from the RGEs.


The Yukawa couplings of the bottom quark ($b$) and the tau lepton ($\tau$)
are proportional to $\tan\beta$ and are thus greatly enhanced
when $\tan\beta$ is large.
In supersymmetric grand unified theories,
the masses of the third generation sfermions
are consequently very sensitive to the value of $\tan\beta$.
As $\tan\beta$ increases, the lighter tau slepton ($\tilde{\tau}_1$)
and the lighter bottom squark ($\tilde{b}_1$) become lighter than
charginos and neutralinos while other sleptons and squarks remain heavy.
Then, $\chi^\pm_1$ and $\chi^0_2$ can dominantly decay into
final states with tau leptons via real or virtual $\tilde{\tau}_1$.

One way to detect $\tau$ leptons is through their one-prong
and three-prong hadronic decays.
The CDF and the D$\emptyset$ collaborations are currently investigating
the efficiencies for detecting these modes
and for implementing a $\tau$ trigger \cite{Taus}.
It has been suggested that the $\tau$ leptons
in the final state may be a promising way
to search for $\chi^\pm_1 \chi^0_2$ production at the Tevatron
if excellent $\tau$ identification becomes feasible
\cite{BCDPT2,Wells,Matchev1,Matchev2}.

Another way of exploiting the $\tau$ signals \cite{Madison},
which we employ in this article,
is to include the soft electrons and muons from leptonic $\tau$ decays
by adopting softer but realistic $p_T$ cuts on the leptons
than conventionally used \cite{Trilepton2}.
We find that this can improve the significance of
the trilepton signal from $\chi^\pm_1 \chi^0_2$ production \cite{Madison}.

After suitable cuts,
there are two major sources of the SM background
\cite{Madison,Trilepton2,Trilepton3,BCDPT2,Matchev1,Matchev2,Baer&Tata,Chanowitz,Ellis}:
(i) $q\bar{q} \to W^* Z^*, W^* \gamma^* \to \ell\nu \ell\bar{\ell}$
or $\ell'\nu' \ell\bar{\ell}$ ($\ell = e$ or $\mu$)
with one or both gauge bosons being virtual\footnote{
If it is not specified,
$W^*$ and $Z^*$ represent real or virtual gauge bosons,
while $\gamma^*$ is a virtual photon.}, and
(ii) $q\bar{q} \to W^* Z^*, W^* \gamma^* \to \ell\nu\tau\bar{\tau}$
or $\tau\nu \ell\bar{\ell}$ and subsequent $\tau$ leptonic decays.
In this article, we substantially improve the background calculations
including effects from virtual $W$ and $Z$ and the contributions
from virtual photons that are missing in ISAJET
and not included in earlier studies
These contributions are unexpectedly important.
We reoptimize the acceptance cuts to reduce the larger background
that resulted.


The experimental measurements of the $b \to s\gamma$ decay rate
by the CLEO \cite{CLEO} and LEP collaborations \cite{LEP}
place constraints on the parameter space
of the minimal supergravity model \cite{bsg}.
It was found that $b \to s\gamma$ disfavors most of
the mSUGRA parameter space
when $\tan\beta \agt 10$ and $\mu < 0$ \cite{bsg}.
Therefore, we concentrate on $\mu > 0$ in our analysis
when $\tan\beta\geq 10$.

In Section II we discuss
the $p\bar{p} \to \chi^\pm_1 \chi^0_2 +X$ cross section
and the decay branching fractions of $\chi^\pm_1$ and $\chi^0_2$.
The acceptance cuts for the signal and background are discussed
in Section III.
We present the trilepton cross section from additional SUSY sources
in Section~IV.
When the sleptons ($\tilde{\ell}$) and the sneutrinos ($\tilde{\nu}$)
are light, they also contribute to the trilepton signal
via production of $\tilde{\ell} \tilde{\ell}$
and $\tilde{\ell} \tilde{\nu}$.
These contributions are at interesting levels when $m_0 \alt 150$ GeV
and $\tan\beta \agt 20$.
The discovery potential of the trilepton search at the upgraded Tevatron
is presented in Section V,
with 3$\sigma$ significance contours for observation or exclusion
and 5$\sigma$ significance contours for discovery.
Section VI discusses the end point reconstruction at the Run III
for invariant mass distribution of $\ell^+\ell^-$ from the $\chi^0_2$ decays.
Our conclusions are given in Section VII.


\section{Associated Production of Chargino and Neutralino}


In hadron collisions the associated production of the lighter chargino
and the second lightest neutralino occurs via quark-antiquark annihilation
in the $s$-channel through a $W$ boson
($q\bar{q'} \to W^{\pm} \to \chi^\pm_1 \chi^0_2$) and
in the $t$ and $u$-channels through squark ($\tilde{q}$) exchanges.
Figure 2 shows the Feynman diagrams of $q\bar{q}' \to \chi^\pm_1 \chi^0_2$.
The $p\bar{p} \to \chi^\pm_1 \chi^0_2 +X$ cross section
depends mainly on masses of the chargino ($m_{\chi^\pm_1}$) and
the neutralino ($m_{\chi^0_2}$).
For squarks much heavier than the gauge bosons,
the $s$-channel $W$-resonance amplitude dominates.
If the squarks are light,
a destructive interference between the $W$ boson and the squark exchange
amplitudes can suppress the cross section by as much as $40\%$,
compared to the $s$-channel contribution alone.
For larger squark masses,
the effect of negative interference is reduced.


Feynman diagrams of the chargino and neutralino decays
into final states of leptons and neutrinos and the LSP are shown
in Fig. 3:
(a) $\chi^\pm_1 \to \ell\nu\chi^0_1 \; {\rm or} \; \tau\nu\chi^0_1$
and
(b) $\chi^0_2 \to \ell^+ \ell^- \chi^0_1 \; {\rm or} \;
\tau^+ \tau^- \chi^0_1$.
Figure 4 presents the branching fractions of $\chi^0_2$
versus $\tan\beta$, with $m_{1/2} = 200$ GeV and several values of $m_0$
for both $\mu > 0$ and $\mu < 0$.
For $\tan\beta \alt 5$,
the branching fractions are sensitive to the sign of $\mu$.

For $\mu > 0$, and $\tan\beta \sim 2$,
we have found that the dominant decays are:
\begin{tabbing}
120 GeV $\alt m_0 \alt$ 170 GeV: AAA \=
 $\chi^0_2 \to \tilde{\ell}_R \ell$ and $\tilde{\tau}_1\tau$
($\chi^0_2 \to \tilde{\nu}_L\nu$ suppressed); \kill
$m_0 \alt$ 50 GeV: \>
$\chi^\pm_1 \to \tilde{\nu}_L\ell$ and $\tilde{\tau}_1\nu$, \\
                  \>
$\chi^0_2 \to \tilde{\ell}_R\ell$, $\tilde{\tau}_1\tau$
         and $\tilde{\nu}_L\nu$  ; \\
 60 GeV $\alt m_0 \alt$ 110 GeV: \>
$\chi^\pm_1 \to \tilde{\tau}_1\nu$, \\
                  \>
$\chi^0_2 \to \tilde{\ell}_R \ell$ and $\tilde{\tau}_1\tau$,
($\chi^0_2 \to \tilde{\nu}_L\nu$ suppressed); \\
120 GeV $\alt m_0 \alt$ 170 GeV: \>
$\chi^\pm_1 \chi^0_2 \to 3 \ell +\notE_T$ via virtual $\tilde{\ell}$; \\
$m_0 \agt 180$ GeV: \>
$\chi^\pm_1, \chi^0_2 \to q\bar{q} \chi^0_1$.
\end{tabbing}

For $\mu < 0$ and $\tan\beta \sim 2$,
we have found that the dominant decays are:
\begin{tabbing}
$m_0 \alt$ 100 GeV: AAA \= $\chi^\pm_1 \to \tilde{\nu}_L\ell$, space \kill
$m_0 \alt$ 100 GeV: \> $\chi^\pm_1 \to \tilde{\nu}_L\ell$, \\
                    \> $\chi^0_2 \to \tilde{\nu}_L\nu$; \\
$m_0 \agt$ 110 GeV: \> $\chi^\pm_1 \to \tilde{\tau}_1\nu$, \\
                    \> $\chi^0_2   \to \chi^0_1 h^0$.
\end{tabbing}

For $m_0 \sim $ 200 GeV, $\chi^0_2$ dominantly decays
(i) into $\tau\bar{\tau}\chi^0_1$ for $25 \alt \tan\beta \alt 40$,
(ii) into $\tau\tilde{\tau}_1$ for $\tan\beta \agt 40$.
For $m_0 \alt 300$ GeV and large $\tan\beta \agt 35$,
both $\tilde{\tau}_1$ and $\tilde{b}_1$ can be lighter than other sfermions,
and $\chi^\pm_1$ and $\chi^0_2$ can decay dominantly into final states
with $\tau$ leptons or $b$ quarks
via virtual or real $\tilde{\tau}_1$ and $\tilde{b}_1$.
For $m_0 \agt 400$ GeV and $5 \alt \tan\beta \alt 40$,
$B(\chi^0_2 \to \tau^+\tau^- \chi^0_1)
\sim B(\chi^0_2 \to e^+e^- \chi^0_1) \sim 2\%$

Figure 5 shows the cross section
$\sigma(p\bar{p} \to \chi^\pm_1 \chi^0_2 \to 3\ell +X)$
at $\sqrt{s} = 2$ TeV,
which is the product
$\sigma(p\bar{p} \to \chi^\pm_1 \chi^0_2 +X) \times
B(\chi^\pm_1 \to \ell \nu \chi^0_1) \times
B(\chi^0_2 \to \ell^+ \ell^- \chi^0_1)$,
versus $\tan\beta$ without acceptance cuts ,
with $m_{1/2} = 200$ GeV and several values of $m_0$
for both $\mu > 0$ and $\mu < 0$.
For $\tan\beta \alt 5$,
the branching fractions are sensitive to the sign of $\mu$.
For $\mu < 0$ and $\tan\beta \sim 2$,
(a) with $m_0 = 100$ GeV,
$B(\chi^0_2 \to \tilde{\nu} \nu) = 0.71$ and
$B(\chi^0_2 \to h^0 \chi^0_1) = 0.19$,
some trileptons are due to $\chi^0_2 \to \tilde{\ell}_R \ell$
and $\tilde{\tau}_1 \tau$;
(b) with $m_0 = 200$ GeV, $B(\chi^0_2 \to h^0 \chi^0_1) = 0.99$
and consequently the trilepton rate drops sharply.
For $m_0 = 100$ GeV and $m_{1/2} = 200$ GeV,
the curves end at $\tan\beta = 28$,
because the region with $\tan\beta \agt$ 28 is theoretically forbidden.

\section{Acceptance Cuts}

In this section we present results from simulations for the trilepton signal
with an event generator and a simple calorimeter
including our acceptance cuts.
The ISAJET 7.40 event generator program \cite{ISAJET}
with the parton distribution functions of CTEQ3L \cite{CTEQ3L}
is employed to calculate the $3\ell +\notE_T$ signal
from all possible sources of SUSY particles.
A calorimeter with segmentation
$\Delta\eta \times \Delta\phi = 0.1 \times (2\pi/24)$
extending to $|\eta| = 4$ is used.
We take the energy resolutions of
$\frac{0.7}{\sqrt{E}}$ for the hadronic calorimeter and
$\frac{0.15}{\sqrt{E}}$ for the electromagnetic calorimeter.
Jets are defined to be hadron clusters with $E_T > 15$ GeV in a cone
with $\Delta R \equiv \sqrt{\Delta\eta^2+\Delta\phi^2} = 0.7$.
Leptons with $p_T > 5$ GeV and within $|\eta_{\ell}| < 2.5$
are considered to be isolated if the hadronic scalar $E_T$
in a cone with $\Delta R = 0.4$ about the lepton is smaller than 2 GeV.

The trilepton signal has dominant physics backgrounds
from production of $W^*V^*$, $V^*V^*$ ($V = Z$ or $\gamma$),
and $t\bar{t}$.
Most backgrounds from the SM processes can be removed
with the following basic cuts:
\begin{enumerate}
\item
We require three isolated leptons in each event\footnote{
The events with 4 isolated leptons are considered as
4-lepton signals in our analysis.}
with $p_T > 5$ GeV and $|\eta_{\ell}| < 2.0$
and that a hadronic scalar $E_T$ smaller than 2 GeV
in a cone with $\Delta R = 0.4$ around the lepton.
This isolation cut removes background from $b\bar{b}$ and $c\bar{c}$ decays.
\item
We require $\notE_T > 25$ GeV in each event to remove backgrounds
from SM processes such as Drell-Yan dilepton production,
where an accompanying jet may fake a lepton.
\item
To reduce the background from $W^*Z^*$ production,
we require that the invariant mass of any opposite-sign dilepton pair
with the same flavor not reconstruct the $Z$ mass:
$|M_{\ell\bar{\ell}} -M_Z| \geq 10$ GeV.
\item
To eliminate the background from $J/\psi$ and $\Upsilon$, and
to reduce the background from $W^*\gamma^*$ production,
we require a minimal value for
the invariant mass of any opposite-sign dilepton pair
with the same flavor: $M_{\ell\bar{\ell}} \geq 12$ GeV.
A more severe $M_{\ell\bar{\ell}}$ cut is imposed later
in Eq.~(\ref{eq:cuts2}).
\end{enumerate}
Our acceptance cuts are chosen to be consistent with the experimental cuts
proposed for Run II \cite{Kamon,Jane} at the Tevatron as follows:
\begin{eqnarray}
p_T(\ell_1) &>& 11 \; {\rm GeV}, \;\; p_T(\ell_2) > 7 \; {\rm GeV}
                                 \;\; p_T(\ell_3) > 5 \; {\rm GeV},
                                 \nonumber \\
|\eta(\ell_1,\ell_2,\ell_3)| &<& 2.0, \nonumber \\
\notE_T &>& 25 \;\; {\rm GeV}, \nonumber \\
M_{\ell\bar{\ell}}       &\geq& 12 \;\; {\rm GeV} \;\; \nonumber \\
|M_{\ell\bar{\ell}}-M_Z| &\geq& 10 \;\; {\rm GeV},
\label{eq:Basic}
\end{eqnarray}
and at least one lepton with $p_T(\ell) > 11$ GeV and $|\eta(\ell)| < 1.0$.

The surviving total background after these cuts
has contributions from four major sources
via quark anti-quark annihilation (Fig. 6):
(i)   production of $e^\pm\nu\mu^+\mu^-$ and $\mu^\pm\nu e^+e^-$
($\ell'\nu'\ell\bar{\ell}$),
(ii)  production of $e^\pm\nu e^+ e^-$ and $\mu^\pm\nu \mu^+\mu^-$
($\ell\nu\ell\bar{\ell}$),
(iii) production of $e^\pm\nu \tau^+\tau^- +\mu^\pm\nu \tau^+\tau^-$
($\ell\nu\tau\bar{\tau}$),
with subsequent $\tau$ leptonic decays, and
(iv) production of $\tau^\pm\nu e^+e^- +\tau^\pm\nu \mu^+\mu^-$
($\tau\nu\ell\bar{\ell}$),
with subsequent $\tau$ leptonic decays.
In addition, there are contributions from the production of
$e\bar{e} \tau\bar{\tau} +\mu\bar{\mu} \tau\bar{\tau}$
($\ell\bar{\ell}\tau\bar{\tau}$),
with one $\tau$ decaying leptonically and another decaying hadronically.
We employed the programs MADGRAPH \cite{Madgraph}
and HELAS \cite{Helas} to evaluate
the background cross section of $p\bar{p} \to 3\ell +\notE_T +X$
for contributions from all these five subprocesses.
The background from $t\bar{t}$ was calculated with ISAJET.

We present invariant mass distribution of the same-flavor lepton pairs
with opposite signs in Fig. 7
for the dominant background from $q\bar{q}' \to \ell'\nu' \ell\bar{\ell}$,
with the basic cuts in Eq. (\ref{eq:Basic}), but without the $Z$ veto.
This background cross section from $W^*\gamma^*$ increases sharply
as the invariant mass becomes smaller for $M_{\ell\bar{\ell}} \alt 30$ GeV.
Therefore, a more stringent dilepton invariant mass cut
than that in Eq. (\ref{eq:Basic}) is necessary
to reduce the background from $W^*\gamma^*$.

Figure 8 shows the transverse mass [$M_T(\ell,\notE_T$] distribution
of the lepton associated with two same-flavor and opposite-sign leptons,
$d\sigma/dM_T(p\bar{p} \to e\nu\mu\bar{\mu} +\mu\nu e\bar{e} +X)$,
from the dominant background
$q\bar{q}' \to e\nu \mu\bar{\mu}+\mu\nu e\bar{e}$
at the upgraded Tevatron with the basic cuts in Eq. (\ref{eq:Basic}).
Also shown are the same distributions of trileptons
[$p_T(\ell_1) \geq p_T(\ell_2) \geq p_T (\ell_3)$]
from the SUSY signal for
$\mu >0$, $\tan\beta =3$, $m_{1/2} = 200$ GeV and $m_0 = 100$ GeV
with the basic cuts.
This figure suggests that a cut on the transverse mass [$M_T(\ell,\notE_T)$]
around $M_W$ can efficiently reduce the backgrounds from $W^*Z^*+W^*\gamma^*$.

To further reduce the background from $W^* Z^*$ and $W^* \gamma^*$,
we require that
\begin{eqnarray}
|M_{\ell\bar{\ell}}-M_Z| &\geq& 15 \;\; {\rm GeV} \;\; (Z \; {\rm veto}),
                         \nonumber \\
M_{\ell\bar{\ell}}       &\geq& 18 \;\; {\rm GeV} \;\; (\gamma \; {\rm veto}),
                         \nonumber \\
M_T(\ell',\notE_T)      &\leq& 65 \;\; {\rm GeV} \;\; {\rm or} \;\;
M_T(\ell',\notE_T)      \geq 85 \;\; {\rm GeV} \;\; (W \; {\rm veto}).
\label{eq:cuts2}
\end{eqnarray}
where $M_{\ell\bar{\ell}}$ is the invariant mass for
any pair of leptons with the same flavor and opposite signs,
and $M_T(\ell',\notE_T)$ is the transverse mass
of the lepton associated with $\ell\bar{\ell}$.

Some $3\ell$ events could be due to $Z+jets$ and $W+jets$.
Since these sources always originate from $b\to c\ell\nu$
followed by $c\to s\ell\nu$,
they can be removed by imposing an angular separation cut
between the isolated leptons,
giving a background consistent with zero.
This angular separation cut causes almost no signal loss.

The transverse momentum ($p_T$) distribution for the three leptons
of the dominant background is shown in Fig. 9
for $p\bar{p} \to e\nu \mu\bar{\mu} +\mu\nu e\bar{e} +X$.
We label the trileptons as $\ell_{1,2,3}$, where $\ell = e$ or $\mu$,
according to the ordering $p_T(\ell_1) > p_T(\ell_2) > p_T(\ell_3)$
of their transverse momenta.
Figure 10 presents the transverse momentum distribution
of the three leptons from the SUSY signal
with $\mu > 0$, $\tan\beta = 10$, $m_{1/2} = 200$ GeV and $m_0 =$ 100 GeV.
The most important lesson we learn from Figure  10 is that
a large number of $\ell_3$'s from the SUSY particle decays
have a $p_T$ less than 5 GeV.
Therefore, it is very important to have a soft $p_T$ acceptance cut
on $\ell_3$ to retain the trilepton events from SUSY sources \cite{Madison}.

The effects of acceptance cuts on the signal and background
are demonstrated in Table~I.
The trileptons are due to $\chi^\pm_1\chi^0_2$ production and
the additional SUSY particle sources that are discussed in the next section.
The cross sections of the signal with $m_{1/2} = 200$ GeV, $m_0 = 100$ GeV,
and several values of $\tan\beta$,
along with $\ell\nu\ell\bar{\ell}$, $t\bar{t}$ and $ZZ$ backgrounds
are presented for four sets of cuts:
(a) Basic Cuts: acceptance cuts in Eq. (\ref{eq:Basic});
(b) Soft Cuts A: acceptance cuts in Eqs. (\ref{eq:Basic})
and (\ref{eq:cuts2});
(c) Soft Cuts B: the same cuts as soft cuts A,
except requiring 18 GeV $\leq M_{\ell\bar{\ell}} \leq$ 75 GeV;
(d) Hard cuts: the same cuts as soft cuts A,
except requiring $M_{\ell\bar{\ell}} \geq 12 \;\; {\rm GeV}$,
and $p_T(\ell_1,\ell_2,\ell_3) >$ 20, 15, and 10 GeV.
We observe that the soft cuts can considerably enhance
the signal significance.
A more strict cut to require $M_{\ell\bar{\ell}} < 75$ GeV as in soft cuts B
can further reduce the backgrounds from $\ell'\nu'\ell\bar{\ell}$
as well as $\ell\nu\ell\bar{\ell}$
with a slight reduction in the trilepton signal for most SUGRA
parameters and might slightly improve the statistical significance.
The reach with each of the soft cuts is qualitatively similar.
For brevity, we will present results with soft cuts A in this article.

In Table II, we present masses of relevant SUSY particles
for four sets of mSUGRA parameters:\footnote{
These cases were selected for the SUGRA study
in the RUN II Workshop on Supersymmetry/Higgs at the Fermilab \cite{Run2}.}
The trilepton signal cross sections and values of statistical significance
for these sets of parameters are presented in Table III
with an integrated luminosity of ${\cal L} = 30$~fb$^{-1}$.
\begin{itemize}
\item Case I:
In this case,
$m_{\tilde{\ell}_R} \sim m_{\tilde{\tau}_1}
< m_{\chi^\pm_1} \sim m_{\chi^0_2}$,
$B(\chi^0_2   \to \tilde{\ell}_R\ell) = 58\%$,
$B(\chi^0_2   \to \tilde{\tau}_1\tau) = 41\%$, and
$B(\chi^\pm_1 \to \tilde{\tau}_1\nu)  = 49\%$,
so a large rate for trilepton events is expected.
The $\chi^\pm_1\chi^0_2$ production contributes about $81\%$
of the total trilepton signal.
The mSUGRA parameters for this case are in the
cosmologically favored region of parameter space
with an appropriate $\chi^0_1$ relic density
for cold dark matter ($\Omega_{\chi^0_1} h^2 = 0.24$)
\cite{Relic,Relic2}.
\item Case II:
This parameter space point has a large value
of $\tan\beta =35$ and $\tilde{\tau}_1$ is much lighter than
$\chi^\pm_1$ and $\chi^0_2$;
$B(\chi^0_2   \to \tilde{\tau}_1\tau) \sim 100\%$, and
$B(\chi^\pm_1 \to \tilde{\tau}_1\nu)  \sim 100\%$.
Here, we anticipate that an inclusive
trilepton signal can be extracted with relatively soft lepton $p_T$ cuts,
since the detected leptons typically come from $\tau$ decays.
\item Case III:
This parameter space point also has a large $\tan\beta$,
but the $A_0$ parameter is chosen so that relatively light
$\tilde{t}_1$, $\tilde{b}_1$ and $\tilde{\tau}_1$ are generated;
$B(\chi^0_2   \to \tilde{\tau}_1\tau) \sim 100\%$, and
$B(\chi^\pm_1 \to \tilde{\tau}_1\nu)  \sim 100\%$.
The trileptons should occur at a similar rate as in Case II.
The rather large $\tilde{t}_1\tilde{t}_1$
production cross section may yield an observable $\tilde{t}_1$ signal.
\item Case IV:
This parameter space choice has a small value of $\tan\beta = 3$ and
the $A_0$ has been chosen such that $\tilde{t}_1$ is light,
while both $\ell_R$ and $\tilde{\tau}_1$ are heavier than $\chi^\pm$;
$B(\chi^0_2   \to \chi^0_1 e\bar{e} +\chi^0_1\mu\bar{\mu}) = 6.6\%$,
and $B(\chi^\pm_1 \to e\nu +\mu\nu)  = 23\%$.
This case could also provide an opportunity
to search for $\tilde{t}_1\tilde{t}_1$ production
where $\tilde{t}_1\to b\chi^\pm_1$
with $\chi^\pm_1\to \ell\nu_\ell\chi^0_1$.
\end{itemize}
Most trileptons in cases I and IV have higher $p_T$
than those in cases II and III, because the latter contain
some secondary leptons from $\tau$ decays.

At Run II with 2 fb$^{-1}$ integrated luminosity,
we expect about 4 events per experiment
from the background cross section of 1.97 fb.
Then the signal cross section must yield
a minimum of 6 signal events for discovery;
the Poisson probability for the SM background to fluctuate
to this level is less than $0.8\%$.
At Run III with ${\cal L} = 30$ fb$^{-1}$,
we would expect about 59 background events;
a $5 \sigma$ signal would be 38 events corresponding to
a signal cross section of 1.28 fb,
and a $3 \sigma$ signal would be 23 events corresponding to
a signal cross section of 0.77 fb.

\section{Additional Sources of Trileptons}

In addition to the associated production of $\chi^\pm_1 \chi^0_2$,
there are other SUSY contributions to trilepton events.
\begin{enumerate}
\item
If the sleptons and sneutrinos are light,
$\tilde{\ell}\tilde{\nu}$ and $\tilde{\ell}\tilde{\ell}$
can make important contribution to the trilepton signal
and $\tilde{\nu}\tilde{\nu}$ can make a small contribution.
\item
When the charginos ($\chi^\pm_{1,2}$),
and the neutralinos ($\chi^0_{2,3,4}$) are not too heavy,
they contribute to the trileptons
via $\chi^0_2\chi^0_2$, $\chi^0_2\chi^0_3$, $\chi^0_3\chi^0_4$,
$\chi^\pm_1\chi^0_3$, $\chi^\pm_2\chi^0_3$,
and $\chi^\pm_2\chi^0_4$ production.
\item
When the gluino ($\tilde{g}$), the squarks ($\tilde{q}$),
and the neutralinos ($\chi^0_{2,3,4}$) are not too heavy,
they also contribute to the trileptons
via the production of $\tilde{g}\chi^0_{2,3}$ and $\tilde{q}\chi^0_{2,3}$.
\item
The production of $\tilde{g}\tilde{g}$ and $\tilde{q}\tilde{q}$
also make small trilepton contributions.
\end{enumerate}

For $m_0 \agt 500$ GeV and $m_{1/2} \alt 300$ GeV,
the associated production of $\chi^\pm_1 \chi^0_2$,
contributes at least 95\% of the trilepton signal.
For $m_0 \alt 150$ and $\tan\beta \agt 20$,
production of $\tilde{\ell}\tilde{\nu}$ and $\tilde{\ell}\tilde{\ell}$
can enhance the trilepton cross section and may yield observable signals
at Run III.
We summarize the contributions to trileptons
from various relevant channels for $\mu > 0$ in Table~IV
and for $\mu < 0$ in Table V.

\section{Discovery Potential at the Tevatron}

The cross sections for the trilepton signal after cuts
are shown in Fig. 11 versus $m_{1/2}$
with $\tan\beta = 3$ and several values of $m_0$
for both $\mu > 0$ and $\mu < 0$.
Figure 12 shows the cross sections of the trilepton signal
and background after cuts versus $m_{1/2}$ with several values of $m_0$
and $\mu > 0$ for $\tan\beta = 10$ and $\tan\beta = 35$.
Also shown are lines for
(i) 6 signal events with ${\cal L} = 2$ fb$^{-1}$ and
(ii) a 5$\sigma$ signal as well as a 3$\sigma$ signal
with ${\cal L} = 30$ fb$^{-1}$.

To assess the overall discovery potential of the upgraded Tevatron,
we present
the 99\% C.L. observation contour at Run II
and the $5\sigma$ discovery contour
as well as the 3$\sigma$ observation contour at Run III
in Fig. 13
for $p\bar{p} \to {\rm SUSY \; particles} \to 3\ell +X$
at $\sqrt{s} = 2$ TeV, with soft acceptance cuts
[Eqs.~(\ref{eq:Basic}) and (\ref{eq:cuts2})],
in the parameter space of $(m_0,m_{1/2})$,
with $\tan\beta =$ 2, for (a) $\mu > 0$ and (b) $\mu < 0$.
All SUSY sources of trileptons are included.
Figure 14 shows
the 99\% C.L. observation contour at Run II
and the $5\sigma$ discovery contour
as well as the 3$\sigma$ observation contour at Run III
for $p\bar{p} \to {\rm SUSY \; particles} \to 3\ell +X$
in the $(m_0,m_{1/2})$ parameter space
for $\tan\beta = 10$ and $\tan\beta = 35$.
We have included all SUSY sources of trileptons.
For 180 GeV $\alt m_0 \alt$ 400 GeV and $10 \alt \tan\beta \alt 40$,
the $\chi^0_2$ decays dominantly into $q\bar{q}\chi^0_1$ and
in these regions it will be difficult to establish a supersymmetry signal.

In Fig. 15, we present the contours of 99\% C.L. observation at Run II
and $5\sigma$ discovery as well as 3$\sigma$ observation at Run III
for $p\bar{p} \to {\rm SUSY \; Particles} \to 3\ell +X$
in the $(m_0,m_{1/2})$ plane for $\tan\beta = 3$
with soft cuts A ($|M_{\ell\bar{\ell}}-M_Z| >$ 15 GeV) and
soft cuts B (18 GeV $\leq M_{\ell\bar{\ell}} \leq$ 75 GeV).
The lighter CP-even Higgs scalar mass ($m_h$)
is sensitive to the value of $\tan\beta$.
Taking $m_{1/2} = 200$ GeV, $m_0 = 100$ GeV, $A_0 = 0$ and $\mu > 0$,
we obtain $m_h = 89.5$ GeV for $\tan\beta = 2$ and
$m_h = 99.3$ GeV for $\tan\beta = 3$.

Also shown in Figs. 13, 14 and 15, are the regions that do not satisfy
the following theoretical requirements:
electroweak symmetry breaking (EWSB),
the correct vacuum for EWSB obtained (tachyon free),
and the lightest neutralino as the lightest SUSY particle (LSP).
The region excluded by the $m_{\chi^+_1} \alt 95$ GeV limit
from the chargino search \cite{LEP2} at LEP 2 is indicated.

We calculated cross sections for the signals and the backgrounds
with tree level amplitudes.
However, we expect that our conclusions and discovery contours will be valid
after QCD radiative corrections are included.
Recent studies found that QCD corrections enhance the signal cross section of
$p\bar{p} \to \chi^\pm_1 \chi^0_2 +X$ by about $10-30\%$
for 70 GeV $\alt m_{\chi^+_1} \alt $ 300 GeV \cite{Tilman}.
QCD corrections also enhance the cross section of the dominant background
$p\bar{p} \to W^*Z^*+W^*\gamma^* +X$ by about 30\% \cite{Ellis,Ohnemus}.

\section{Mass Reconstruction}

If the two-body decay
$\chi_2^0 \to \tilde{\ell}_R \ell \to \ell^+\ell^- +\chi^0_1$
is kinematically allowed and a large integrated luminosity is accumulated,
it may be possible to test a predicted mass relation \cite{LHC}
among $m_{\chi^0_2}$, $m_{\tilde{l}_R}$ and $m_{\chi^0_1}$.
To demonstrate this interesting possibility, we consider
the following parameters: $m_{1/2} = 200$ GeV, $m_0 = 100$ GeV,
$A_0 = 0$, $\tan\beta = 3$ and $\mu > 0$.
We evaluate masses and couplings of SUSY particles at the weak scale
with renormalization group equations and obtain
$m_{\chi^0_2} = 143$ GeV, $m_{\tilde{e}_R} = 133$ GeV
$m_{\chi^0_1} = 76.0$ and $\mu = 312$.
The corresponding trilepton cross section after cuts
$\sigma(p\bar{p} \to {\rm SUSY \; particles} \to 3\ell+\notE +X) =$ 8.6 fb,
gives a promising signal with 258 events for ${\cal L} = 30$ fb$^{-1}$.
The $\chi^\pm_1\chi^0_2$ production contributes about $81\%$
of the total trilepton signal.

We consider the subtracted dilepton invariant mass distribution \cite{LHC}
defined as
\begin{eqnarray}
\left.{d\sigma\over dM}\right\vert_{\rm ll} =
\left.{d\sigma\over dM}\right\vert_{e^+e^-}
+\left.{d\sigma\over dM}\right\vert_{\mu^+\mu^-}
-\left.{d\sigma\over dM}\right\vert_{e^+\mu^-}
-\left.{d\sigma\over dM}\right\vert_{e^-\mu^+}\,.
\label{eq:xmll}
\end{eqnarray}
The subtractions remove the lepton pairs
with one lepton coming from $\chi^\pm_1$ and another coming from $\chi^0_2$.
This mass distribution has a sharp edge (endpoint)
that appears near the kinematic limit for this decay sequence, i.e.,
\begin{eqnarray}
M_{\ell\bar{\ell}}^{\rm MAX} = M_{\chi_2^0}
\sqrt{1-{M_{\tilde\ell}^2 \over M_{\chi_2^0}^2}}
\sqrt{1-{M_{\chi^0_1}^2 \over M_{\tilde\ell}^2}}
\approx 45 \,\, {\rm GeV}.
\end{eqnarray}
%

Figure 16 shows the subtracted invariant mass distribution for two leptons
with opposite signs ($\ell^+\ell^-$)
from $p\bar{p} \to {\rm SUSY \; particles} \to 3\ell +X$
with $\mu > 0$, $\tan\beta =$ 3, $m_{1/2} = 200$ GeV, and $m_0 =$ 100 GeV.
This distribution may allow a test of the mSUGRA mass relations
in this optimal case with a high cross section, provided that a large
luminosity accumulation (${\cal L} \agt 10$ fb$^{-1}$) is obtained.

\section{Conclusions}

In most of the mSUGRA parameter space,
$\chi^\pm_1\chi^0_2$ production is the dominant source of trileptons.
For $m_0 \alt 150$ and $\tan\beta \agt 20$,
production of $\tilde{\ell}\tilde{\nu}$ and $\tilde{\ell}\tilde{\ell}$
can enhance the trilepton signal and may yield observable rates at Run III
in regions of parameter space that are otherwise inaccessible.

In some regions of the mSUGRA parameter space,
the $\chi^\pm_1$ and the $\chi^0_2$ decay
dominantly to final states with $\tau$ leptons.
The subsequent leptonic decays of these $\tau$ leptons
contribute importantly to the trilepton signal
from $\chi^\pm_1 \chi^0_2$ associated production.
With soft but realistic lepton $p_T$ acceptance cuts,
these $\tau \to \ell$ contributions
can substantially enhance the statistical significance
of trilepton signal, compared to that with hard cuts.
The branching fractions of $\chi^\pm_1$ and $\chi^0_2$ decays
into $\tau$ leptons are dominant when the universal scalar mass $m_0$
is less than about 200 GeV and/or $\tan\beta \agt 40$.

The Tevatron trilepton searches are most sensitive to the region
of mSUGRA parameter space with $m_0 \alt 100$ GeV and $\tan\beta \alt$ 10.
The discovery possibilities of the upgraded Tevatron for $\mu > 0$
are summarized in the following:
\begin{itemize}
\item For $m_0 \sim 100$ GeV and $\tan\beta\sim 2$,
the trilepton signal should be detectable
at the Run II if $m_{1/2} \alt$ 240 GeV ($m_{\chi^\pm_1} \alt 177$ GeV), and
at the Run III if $m_{1/2} \alt$ 260 GeV ($m_{\chi^\pm_1} \alt 195$ GeV).
\item For $m_0 \sim 150$ GeV and $\tan\beta\sim 35$,
the trilepton signal should be detectable
at the Run III if $m_{1/2} \alt$ 170 GeV ($m_{\chi^\pm_1} \alt 122$ GeV).
\item For $m_0 \agt 600$ GeV and $\tan\beta\sim 35$,
the trilepton signal should be detectable
at the Run III if $m_{1/2} \alt$ 170 GeV ($m_{\chi^\pm_1} \alt 130$ GeV).
\end{itemize}

It might be difficult to establish a trilepton signal
for 180 GeV $\alt m_0 \alt$ 400 GeV and $3 \alt \tan\beta \alt 35$,
because for these parameters $\chi^\pm_1$ and $\chi^0_2$
dominantly decay into  $q\bar{q}'\chi^0_1$,
and the leptonic decays of $\chi^0_2$ is suppressed.
However, the important lesson is that the experiments at the Tevatron
may probe a substantial region not accessible at LEP 2.


\section*{Acknowledgments}

VB thanks the Fermilab theory group for kind hospitality and support
as a Frontier Fellow.
We thank Teruki Kamon and Xerxes Tata for suggestions on the manuscript.
We are grateful to Howie Baer, Regina Demina, Tao Han,
Wai-Yee Keung, Jane Nachtman, Frank Paige,
and Dieter Zeppenfeld for beneficial discussions.
This research was supported in part by the U.S. Department of Energy
under Grants No. DE-FG02-95ER40896
and in part by the University of Wisconsin Research Committee
with funds granted by the Wisconsin Alumni Research Foundation.


%



\begin{table}[htb]
\begin{center}
\caption[]{
The cross section of
$p\bar{p} \to {\rm SUSY \;\; particles} \to 3\ell +X$ in fb
versus $\tan\beta$ for $m_{1/2} = 200$ GeV and $m_0 = 100$ GeV
along with the trilepton cross sections of the SM backgrounds (BG)
and values of statistical significance
($N_S \equiv S/\sqrt{B}$, $S =$ number of signal events,
and $B =$ number of background events)
for an integrated luminosity of ${\cal L} = 30$~fb$^{-1}$,
at the upgraded Tevatron with six sets of cuts:
(a) Basic Cuts: cuts in Eq. (\ref{eq:Basic});
(b) Soft Cuts A:
cuts in Eqs. (\ref{eq:Basic}) and (\ref{eq:cuts2});
(c) Soft Cuts B: the same cuts as soft cuts A1,
except requiring 18 GeV $\leq M_{\ell\bar{\ell}} \leq$ 75 GeV;
(d) Hard Cuts: the same cuts as soft cuts A,
except requiring $M_{\ell\bar{\ell}} \geq 12 \;\; {\rm GeV}$,
and $p_T(\ell_1,\ell_2,\ell_3) >$ 20, 15, and 10 GeV.
}

\medskip

\begin{tabular}{lccccc}
$\tan\beta$ $\backslash$ Cuts
   & Basic & Soft A & Soft B & Hard \\
\hline
3  &  12.8 & 8.82 & 8.41 & 4.04 \\
10 &  3.49 & 2.57 & 2.43 & 1.13 \\
20 &  1.18 & 0.90 & 0.79 & 0.34 \\
25 &  0.66 & 0.50 & 0.43 & 0.20 \\
\hline
SM BG \\
\hline
$\ell'\nu'\ell\bar{\ell}$ & 2.63 & 0.72 & 0.60 & 0.32 \\
$\ell \nu \ell\bar{\ell}$ & 2.09 & 0.41 & 0.30 & 0.20 \\
$\ell \nu \tau\bar{\tau}$ & 0.60 & 0.45 & 0.41 & 0.22 \\
$\tau \nu \ell\bar{\ell}$ & 0.37 & 0.20 & 0.13 & 0.11 \\
$\ell \ell\tau\bar{\tau}$ & 0.12 & 0.08 & 0.06 & 0.04 \\
$t\bar{t}$                & 0.14 & 0.11 & 0.06 & 0.009 \\
Total BG                  & 5.95 & 1.97 & 1.56 & 0.90 \\
\hline
$\tan\beta$ $\backslash$ $N_S$ \\
\hline
3  &  28.7 & 34.4 & 36.9 & 23.3  \\
10 &   7.8 & 10.0 & 10.7 &  6.5  \\
20 &   2.6 &  3.5 &  3.5 &  2.0  \\
25 &   1.5 &  1.9 &  1.9 &  1.2
\end{tabular}
\end{center}
\end{table}


\begin{table}
\caption[]{
Masses of relevant SUSY particles for four mSUGRA cases
along with the value of the Higgs mixing parameter $\mu > 0$.
}
\medskip
\begin{tabular}{crrrr}
Parameters $\backslash$ Cases & I & II & III & IV \\
$m_0$       & 100 & 140  & 200  & 250  \\
$m_{1/2}$   & 200 & 175  & 140  & 150  \\
$A_0$       & 0   & 0    & -500 & -600 \\
$\tan\beta$ & 3   & 35   & 35   & 3    \\
\hline
Masses of relevant SUSY particles \\
\hline
The $\mu$ parameter  & 312 & 241 & 286 & 369  \\
$m_{\tilde{g}}$      & 508 & 455 & 375 & 403  \\
$m_{\tilde{u}_L}$    & 457 & 417 & 375 & 420  \\
$m_{\tilde{d}_L}$    & 463 & 424 & 383 & 426  \\
$m_{\tilde{u}_R}\sim m_{\tilde{d}_R}$ & 440 & 406 & 367 & 413  \\
$m_{\tilde{t}_1}$    & 306 & 297 & 153 & 134  \\
$m_{\tilde{b}_1}$    & 418 & 329 & 213 & 346  \\
$m_{\chi^\pm_1}$     & 141 & 126 & 106 & 109  \\
$m_{\chi^0_1}$       &  76 &  69 &  56 &  57  \\
$m_{\chi^0_2}$       & 143 & 127 & 107 & 111  \\
$m_{\tilde{\ell}_R}$ & 133 & 162 & 212 & 260  \\
$m_{\tilde{\tau}_1}$ & 132 & 104 &  88 & 257  \\
$m_{\tilde{\ell}_L}$ & 180 & 194 & 229 & 275  \\
$m_{\tilde{\nu}_L}$  & 165 & 177 & 214 & 266
\end{tabular}
\end{table}


\begin{table}
\caption[]{
The cross section (in fb) of
$p\bar{p} \to {\rm SUSY \; particles} \to 3\ell +X$ at $\sqrt{s} = 2$ TeV
for the four mSUGRA cases described in Table II with contributions
from various SUSY channels.
For each mSUGRA case, the statistical significance $N_S \equiv S/\sqrt{B}$,
$S (B)=$ number of signal (background) events,
is presented for an integrated luminosity ${\cal L} = 30$ fb$^{-1}$
and the four sets of acceptance cuts described in Table I.
}
\medskip
\begin{tabular}{crrrrc}
Acceptance Cuts $\backslash$ Cases & I & II & III & IV & SM Background \\
\hline
Cross Section \\
Basic Cuts   &  12.8 & 1.58 & 1.94 & 4.13 & 5.95 \\
Soft  Cuts A &  8.82 & 1.21 & 1.32 & 3.06 & 1.97 \\
Soft  Cuts B &  8.41 & 0.97 & 1.18 & 2.97 & 1.56 \\
Hard  Cuts   &  4.04 & 0.47 & 0.30 & 1.58 & 0.90 \\
\hline
Statistical Significance: $N_S \equiv S/\sqrt{B}$ \\
\hline
Basic Cuts   &  28.7 & 3.5 & 4.4 &  9.3 &  \\
Soft  Cuts A &  34.4 & 4.7 & 5.2 & 12.0 &  \\
Soft  Cuts B &  36.9 & 4.3 & 5.2 & 13.0 &  \\
Hard  Cuts   &  23.3 & 2.7 & 1.7 &  8.8 &
\end{tabular}
\end{table}


\begin{table}
\caption[]{
The cross section of $p\bar{p} \to 3\ell +X$ in fb versus $\tan\beta$
with contributions from various relevant SUSY channels
at $\sqrt{s} = 2$ TeV with the acceptance cuts described in
Eqs. (\ref{eq:Basic}) and (\ref{eq:cuts2})
for $\mu > 0$, $m_{1/2} = 200$ GeV, $\tan\beta= 2, 10, 20$
and 35 (25 for $m_0 = 100$ GeV).
}
\medskip
\begin{tabular}{crrrc}
Channel $\backslash \tan\beta$ & 2 & 10 & 20 & 35(25) \\
\tableline
(i) $m_0 = 100$ GeV & & & \\
Total
  & 9.58 & 2.57 & 0.90 & 0.50 \\
$\chi^\pm_1\chi^0_2$
  & 7.86 & 1.74 & 0.40 & 0.13 \\
$\tilde{\ell}\tilde{\nu}$
  & 0.68 & 0.32 & 0.18 & 0.10  \\
$\tilde{\ell}\tilde{\ell}$
  & 0.35 & 0.16 & 0.15 & 0.13 \\
$\chi^0_2\chi^0_2$, $\chi^0_2\chi^0_3$, $\chi^0_3\chi^0_4$,
  & 0.30 & 0.12 & 0.05 & 0.05 \\
$\chi^\pm_1\chi^0_{3,4}$, $\chi^\pm_2\chi^0_{3,4}$
$\chi^\pm_1\chi^\mp_2$, $\chi^\pm_2\chi^\mp_2$
  & 0.04 & 0.08 & 0.06 & 0.05 \\
$\tilde{g}\chi^0_{2,3}$,
$\tilde{q}\chi^0_{2,3}$,$\tilde{g}\tilde{g}$,
$\tilde{q}\tilde{q}$,$\tilde{\nu}\tilde{\nu}$
  & 0.35 & 0.15 & 0.06 & 0.04 \\
\tableline
(ii) $m_0 = 200$ GeV & & & \\
Total
  & 2.11 & 0.23 & 0.25 & 0.31 \\
$\chi^\pm_1\chi^0_2$
  & 1.92 & 0.16 & 0.17 & 0.19 \\
$\tilde{\ell}\tilde{\nu}$
  & 0.06 & 0.02 & 0.02 & 0.03  \\
$\tilde{\ell}\tilde{\ell}$
  & 0.03 & --   & 0.01 & 0.01 \\
$\chi^0_2\chi^0_2$, $\chi^0_2\chi^0_3$, $\chi^0_3\chi^0_4$,
  & 0.02 & --   & 0.01 & 0.02 \\
$\chi^\pm_1\chi^0_{3,4}$, $\chi^\pm_2\chi^0_{3,4}$
$\chi^\pm_1\chi^\mp_2$, $\chi^\pm_2\chi^\mp_2$
  & 0.01 & 0.02 & 0.02 & 0.02 \\
$\tilde{g}\chi^0_{2,3}$,
$\tilde{q}\chi^0_{2,3}$,$\tilde{g}\tilde{g}$,
$\tilde{q}\tilde{q}$,$\tilde{\nu}\tilde{\nu}$
 &  0.07 & 0.03 & 0.02 & 0.04 \\
\tableline
(iii) $m_0 = 500$ GeV & & &  \\
Total
  & 0.27 & 0.48 & 0.46 & 0.42 \\
$\chi^\pm_1\chi^0_2$
  & 0.26 & 0.46 & 0.45 & 0.41 \\
$\chi^0_2\chi^0_2$, $\chi^0_2\chi^0_3$, $\chi^0_3\chi^0_4$,
  & --   & 0.01 & --   & --   \\
$\chi^\pm_1\chi^0_{3,4}$, $\chi^\pm_2\chi^0_{3,4}$
$\chi^\pm_1\chi^\mp_2$, $\chi^\pm_2\chi^\mp_2$
  & 0.01 & 0.01 & 0.01 & 0.01   \\
$\tilde{g}\chi^0_{2,3}$,
$\tilde{q}\chi^0_{2,3}$,$\tilde{g}\tilde{g}$,
$\tilde{q}\tilde{q}$,$\tilde{\nu}\tilde{\nu}$
  & --  & --    & --  & --
\end{tabular}
\end{table}


\begin{table}
\caption[]{
The cross section (in fb) of $p\bar{p} \to 3\ell +X$ at $\sqrt{s} = 2$ TeV
with contributions from various SUSY channels
and the acceptance cuts described in
Eqs. (\ref{eq:Basic}) and (\ref{eq:cuts2}),
for $\mu < 0$, $m_{1/2} = 160$ GeV, $\tan\beta= 2$
and several choices of $m_0$.
}
\medskip
\begin{tabular}{crrrc}
Channel $\backslash m_0$ (GeV)& 100 & 200 & 500 & 1000 \\
\tableline
Total
  & 4.91 & 3.24 & 1.12 & 0.89 \\
$\chi^\pm_1\chi^0_2$
  & 3.74 & 2.62 & 1.07 & 0.88 \\
$\tilde{\ell}\tilde{\nu}$
  & 0.12 & 0.15 & --   & -- \\
$\tilde{\ell}\tilde{\ell}$
  & 0.12 & 0.03 & --   & -- \\
$\chi^0_2\chi^0_2$, $\chi^0_2\chi^0_3$, $\chi^0_3\chi^0_4$,
  & 0.20 & 0.09 & 0.01 & -- \\
$\chi^\pm_1\chi^0_{3,4}$, $\chi^\pm_2\chi^0_{3,4}$
$\chi^\pm_1\chi^\mp_2$, $\chi^\pm_2\chi^\mp_2$
  & 0.04 & 0.02 & --   & -- \\
$\tilde{g}\chi^0_{2,3}$,
$\tilde{q}\chi^0_{2,3}$,$\tilde{g}\tilde{g}$,
$\tilde{q}\tilde{q}$,$\tilde{\nu}\tilde{\nu}$
 &  0.69 & 0.32 & 0.04 & 0.01
\end{tabular}
\end{table}


\begin{figure}
\centering\leavevmode
\epsfxsize=6in\epsffile{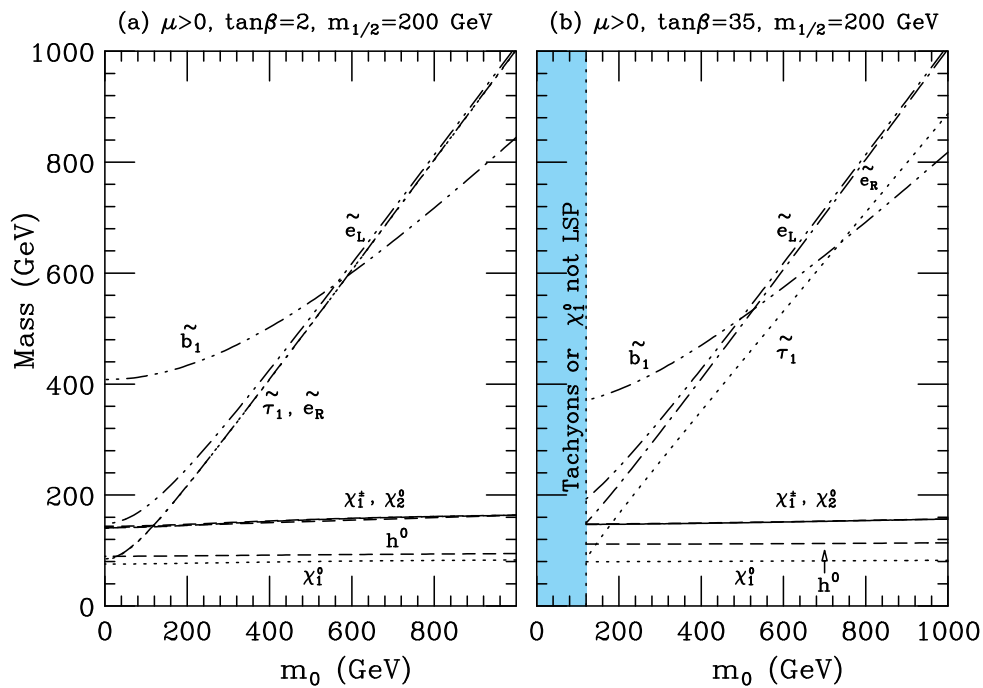}

\bigskip
\caption[]{
Masses of $\chi^0_1$, $\chi^0_2$, $\chi^\pm_1$,
$\tilde{e}_L$, $\tilde{e}_R$, $\tilde{\tau}_1$ and $\tilde{b}_1$
at the $M_Z$ mass scale
and mass of $h^0$ at the mass scale
$Q = \sqrt{  m_{\tilde{t}_L} m_{\tilde{t}_R} }$,
versus $m_0$,
with $M_{\rm SUSY} =$ 1 TeV, $m_{1/2} = 200$ GeV and $\mu > 0$ for
(a)~$\tan\beta = 2$ and (b)~$\tan\beta = 35$.
The shaded regions are excluded
by theoretical requirements (tachyons and LSP).
\label{fig:mass}
}\end{figure}

\begin{figure}
\centering\leavevmode
\epsfxsize=6in\epsffile{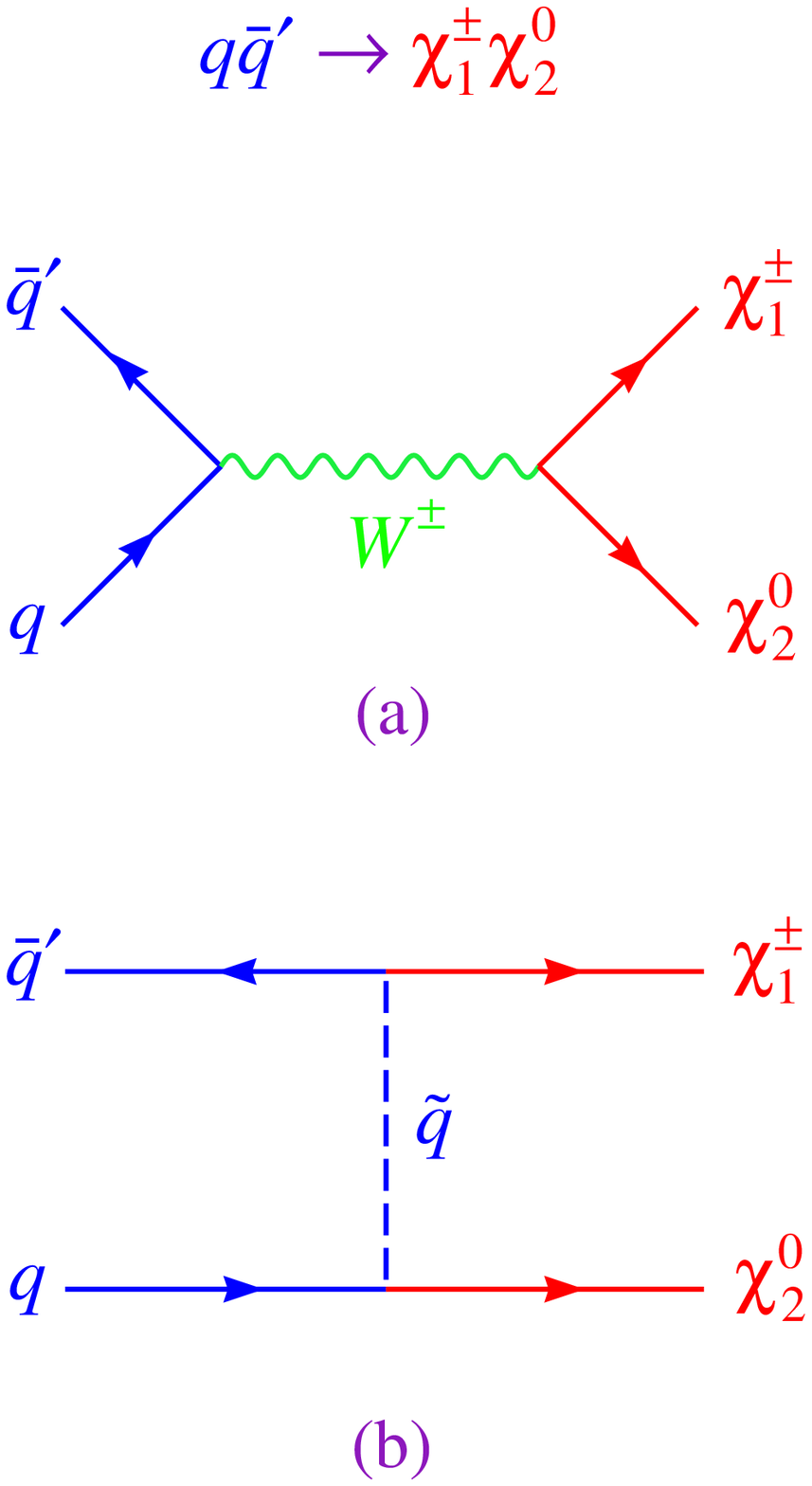}

\bigskip
\caption[]{
Feynman diagrams of $q\bar{q}' \to \chi^\pm_1 \chi^0_2$.
\label{fig:Feynman1}
}\end{figure}

\begin{figure}
\centering\leavevmode
\epsfxsize=6in\epsffile{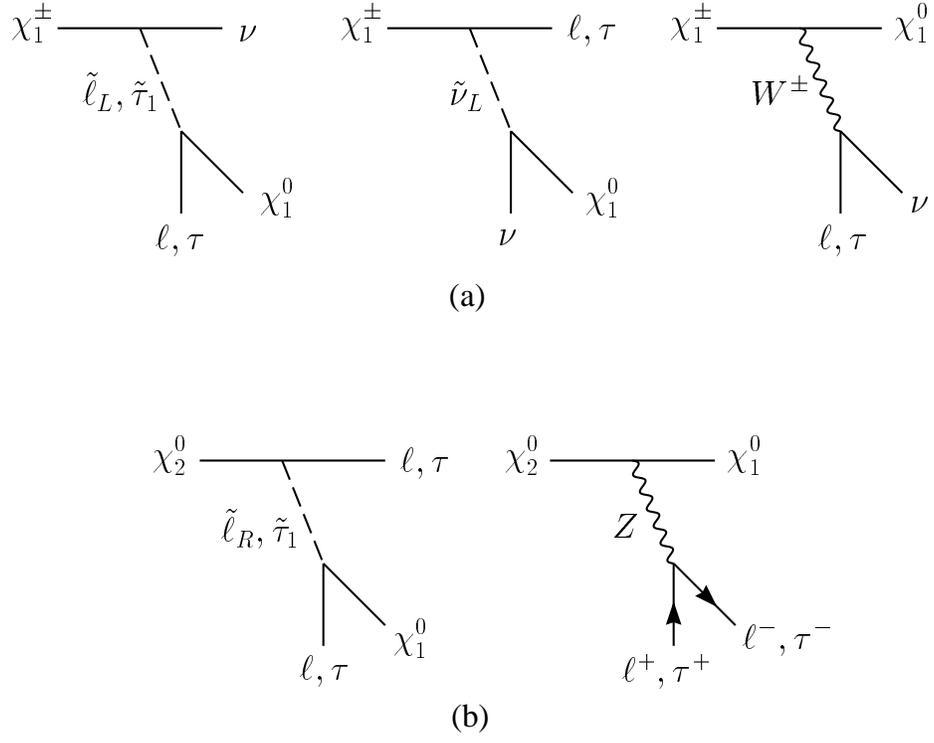}

\bigskip
\caption[]{
Feynman diagrams of
(a) $\chi^\pm_1 \to \ell\nu\chi^0_1 \;\; {\rm or} \;\; \tau\nu\chi^0_1$
and
(b) $\chi^0_2 \to \ell^+ \ell^- \chi^0_1 \;\; {\rm or} \;\;
\tau^+ \tau^- \chi^0_1$.
\label{fig:Decay}
}\end{figure}

\begin{figure}
\centering\leavevmode
\epsfxsize=5.75in\epsffile{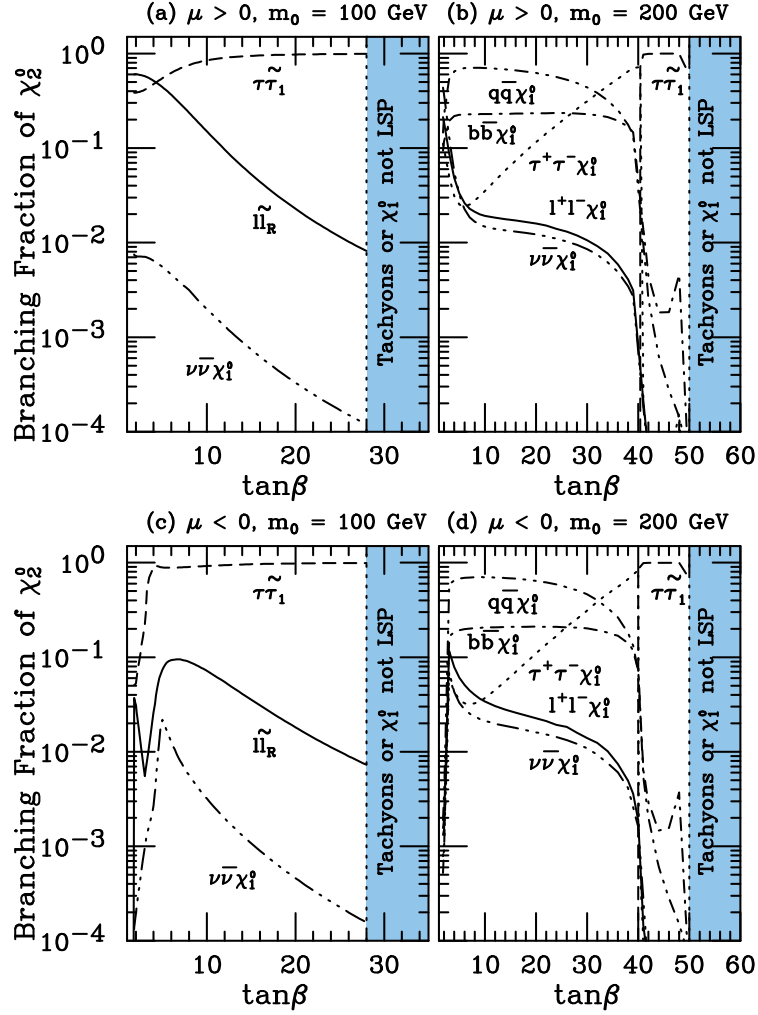}

\bigskip
\caption[]{
Branching fractions of $\chi^0_2$ decays into various channels
versus $\tan\beta$ with $m_{1/2} =$ 200 GeV, for
(a) $\mu > 0$ and $m_0 = 100$ GeV, (b) $\mu > 0$ and $m_0 = 200$ GeV,
(c) $\mu < 0$ and $m_0 = 100$ GeV, and (d) $\mu < 0$ and $m_0 = 200$ GeV.
\label{fig:bfx2}
}\end{figure}

\begin{figure}
\centering\leavevmode
\epsfxsize=6in\epsffile{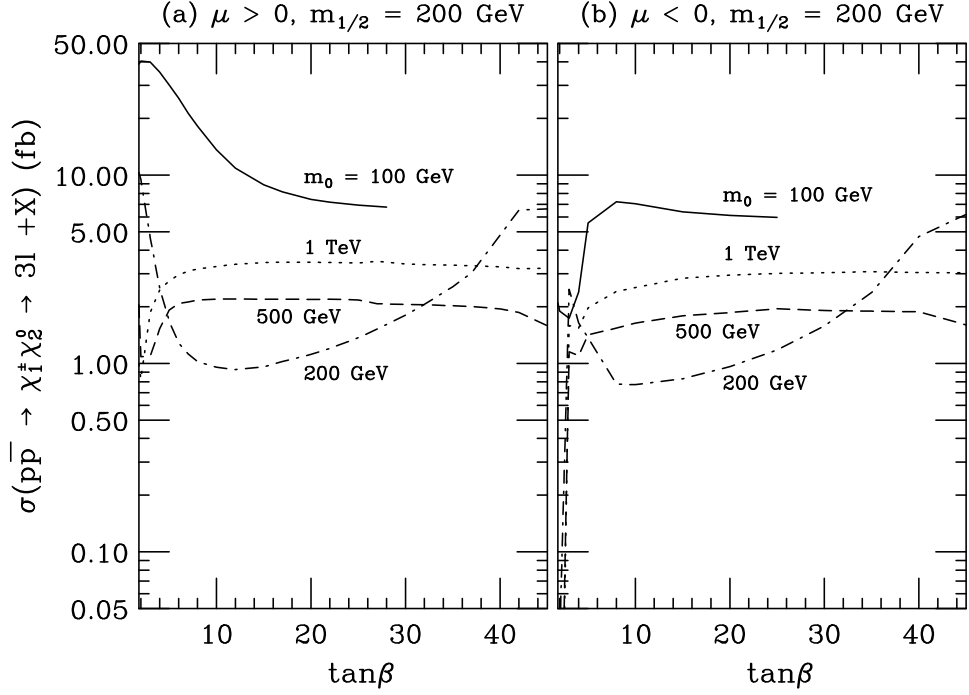}

\bigskip
\caption[]{
Cross section of $p\bar{p} \to \chi^\pm_1 \chi^0_2 \to 3\ell +X$
at $\sqrt{s} = 2$ TeV without cuts versus $\tan\beta$,
with $m_{1/2} = 200$ GeV and several values of $m_0$
for (a) $\mu > 0$ and (b) $\mu < 0$.
For $m_0 = 100$ GeV, the curves end at $\tan\beta = 28$,
because the region with $\tan\beta \agt$ 28 is theoretically forbidden.
\label{fig:xmu}
}\end{figure}

\begin{figure}
\centering\leavevmode
\epsfxsize=6in\epsffile{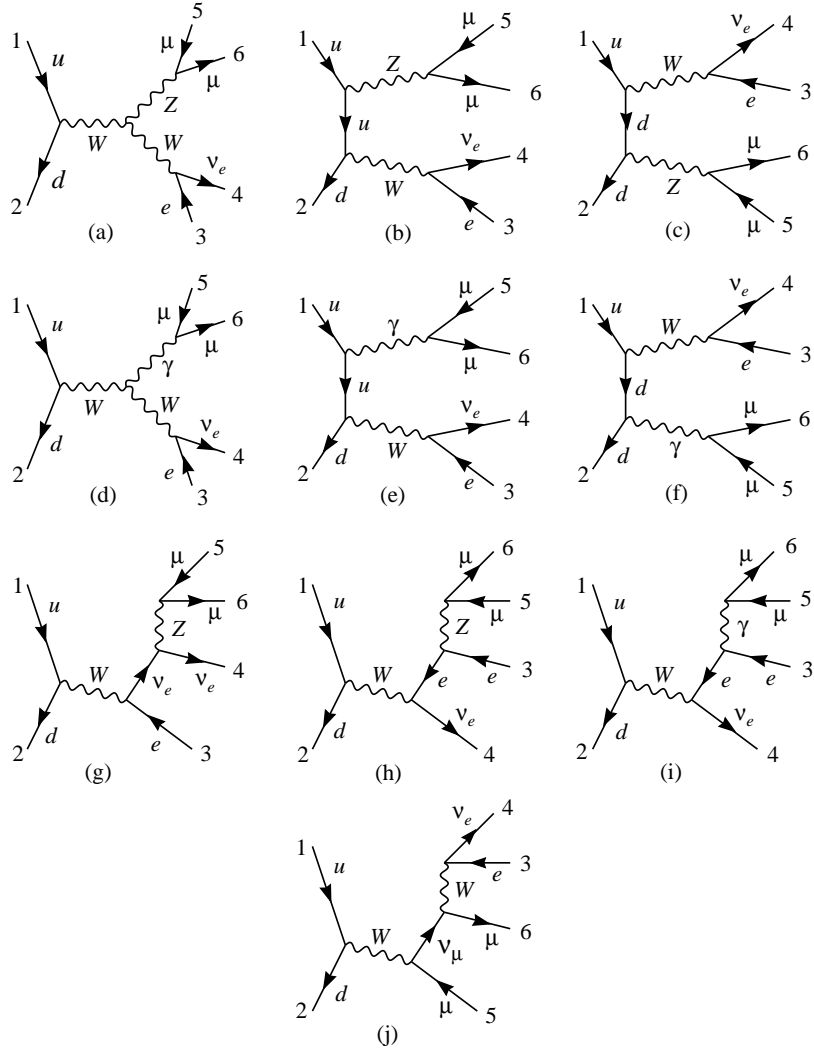}

\bigskip
\caption[]{
Feynman diagrams of $u\bar{d} \to e^+\nu_e \mu^+\mu^-$.
\label{fig:Feynman2}
}\end{figure}

\begin{figure}
\centering\leavevmode
\epsfxsize=6in\epsffile{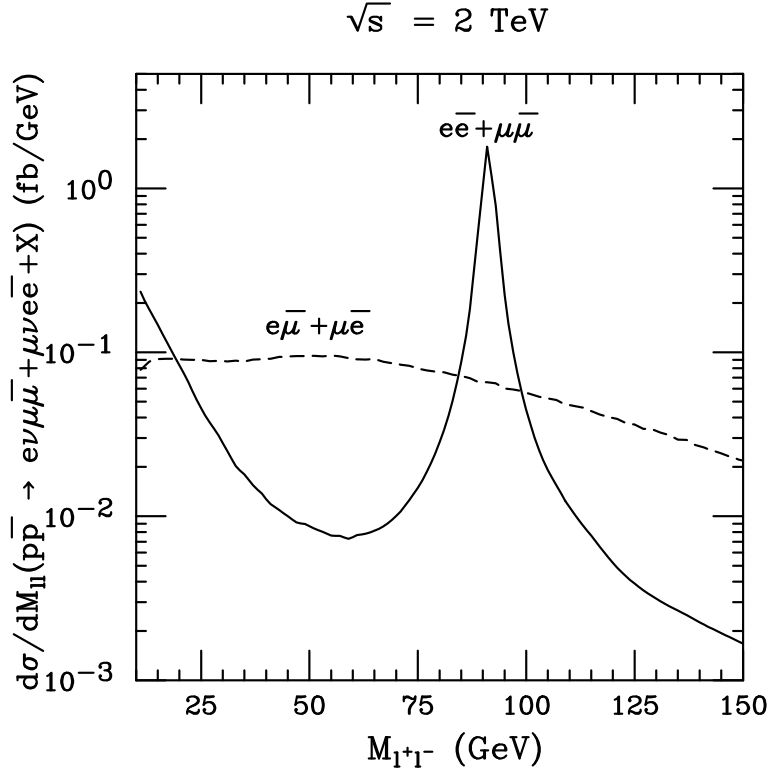}

\bigskip
\caption[]{
Invariant mass ($M_{\ell\bar{\ell}}$) distribution of the lepton pairs
with the same flavor and opposite sign
[$d\sigma/dM_{\ell\bar{\ell}}
(p\bar{p} \to e\nu\mu\bar{\mu} +\mu\nu e \bar{e} +X)$],
for the dominant background from $q\bar{q}' \to \ell'\nu' \ell^+\ell^-$,
at $\sqrt{s} = 2$ TeV, with the basic cuts in Eq. (\ref{eq:Basic}),
but without $Z$-veto.
Also shown is the invariant mass distribution of $e\bar{\mu}+\mu\bar{e}$
with opposite signs.
\label{fig:XMBG}
}\end{figure}
%

\begin{figure}
\centering\leavevmode
\epsfxsize=6in\epsffile{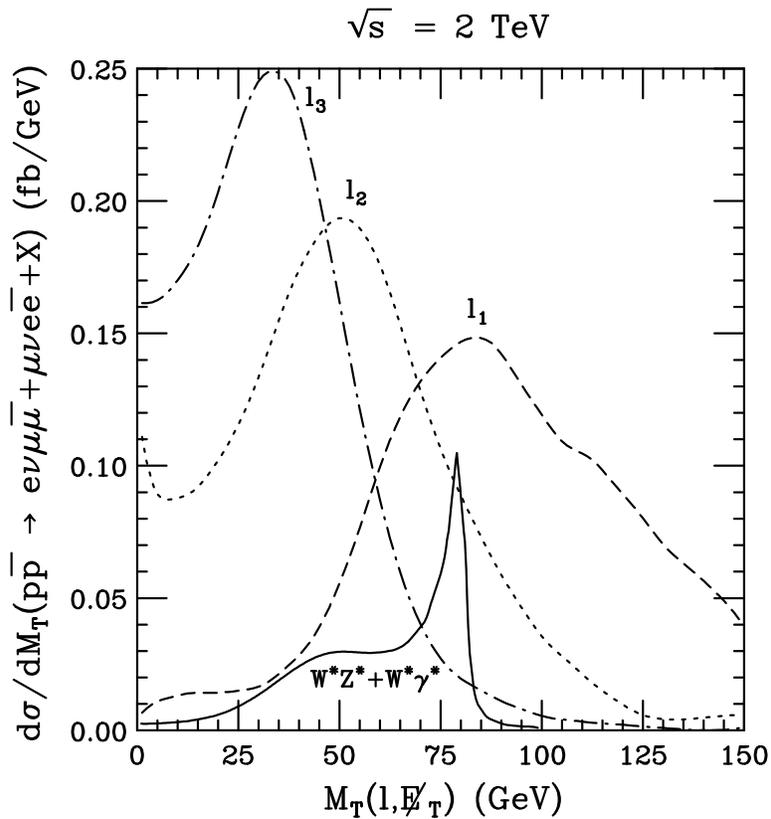}

\bigskip
\caption[]{
Transverse mass [$M_T(\ell,\notE_T$)] distribution of the lepton
associated with two same-flavor and opposite-sign leptons
from the dominant background
$q\bar{q}' \to e\nu \mu\bar{\mu}+\mu\nu e\bar{e}$
with the basic cuts in Eq. (\ref{eq:Basic}).
Also shown are the same distributions of trileptons
[$p_T(\ell_1) \geq p_T(\ell_2) \geq p_T (\ell_3)$]
from the SUSY signal with the basic cuts for
$\mu >0$, $\tan\beta =3$, $m_{1/2} = 200$ GeV and $m_0 = 100$ GeV.
\label{fig:MTen}
}\end{figure}
%

\begin{figure}
\centering\leavevmode
\epsfxsize=6in\epsffile{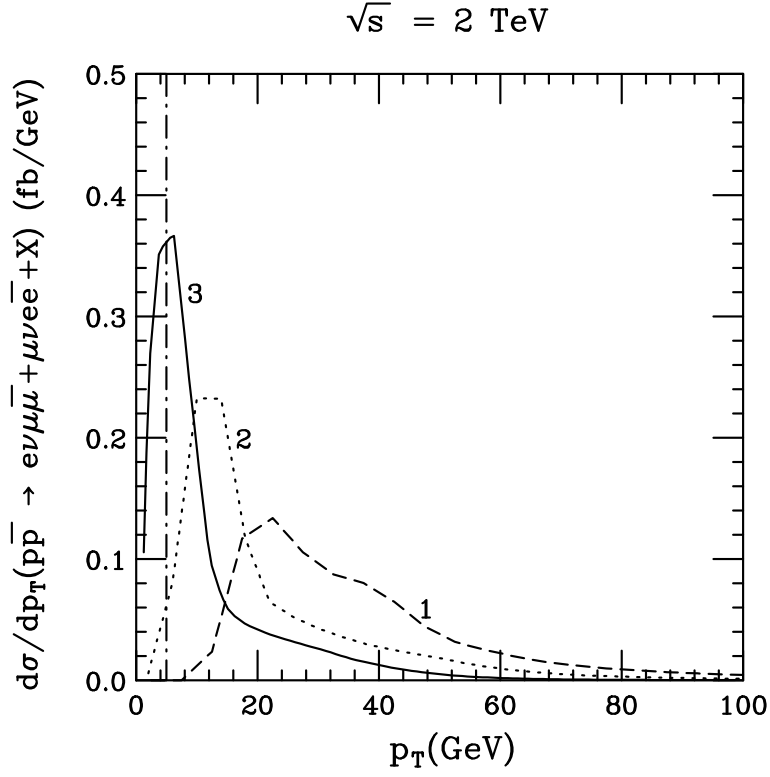}

\bigskip
\caption[]{
Transverse momentum ($p_T$) distribution of the trileptons
from $q\bar{q}' \to \ell' \nu \ell^+\ell^-$ at the upgraded Tevatron
[$d\sigma/dp_T
(p\bar{p} \to e\nu\mu\bar{\mu} +\mu\nu e \bar{e} +X)$],
with the basic cuts in Eq. (\ref{eq:Basic}),
for the three leptons with
$p_T(\ell_1) \geq p_T(\ell_2) \geq p_T (\ell_3) \geq 1$ GeV.
\label{fig:ptenmm}
}\end{figure}
%

\begin{figure}
\centering\leavevmode
\epsfxsize=6in\epsffile{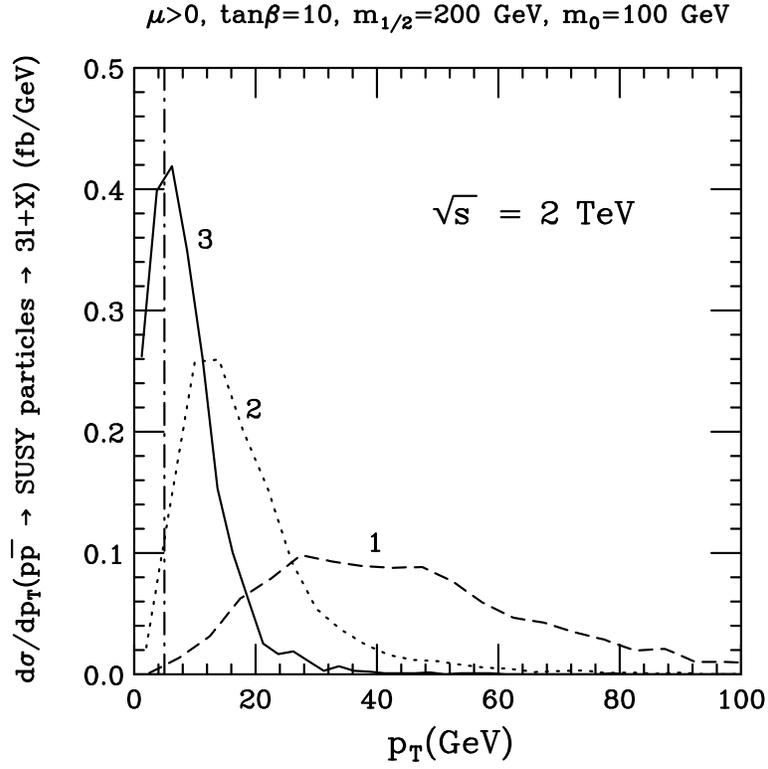}

\bigskip
\caption[]{
Transverse momentum distribution of
$p\bar{p} \to {\rm SUSY \; particles} \to 3\ell +X$ at $\sqrt{s} = 2$ TeV
with the basic cuts in Eq. (\ref{eq:Basic})
and $p_T(\ell_1) \geq p_T(\ell_2) \geq p_T (\ell_3) \geq 1$ GeV,
for $\mu > 0$, $\tan\beta = 10$, $m_{1/2} = 200$ GeV and $m_0 =$ 100 GeV
\label{fig:pt3l}
}\end{figure}
%

\begin{figure}
\centering\leavevmode
\epsfxsize=6in\epsffile{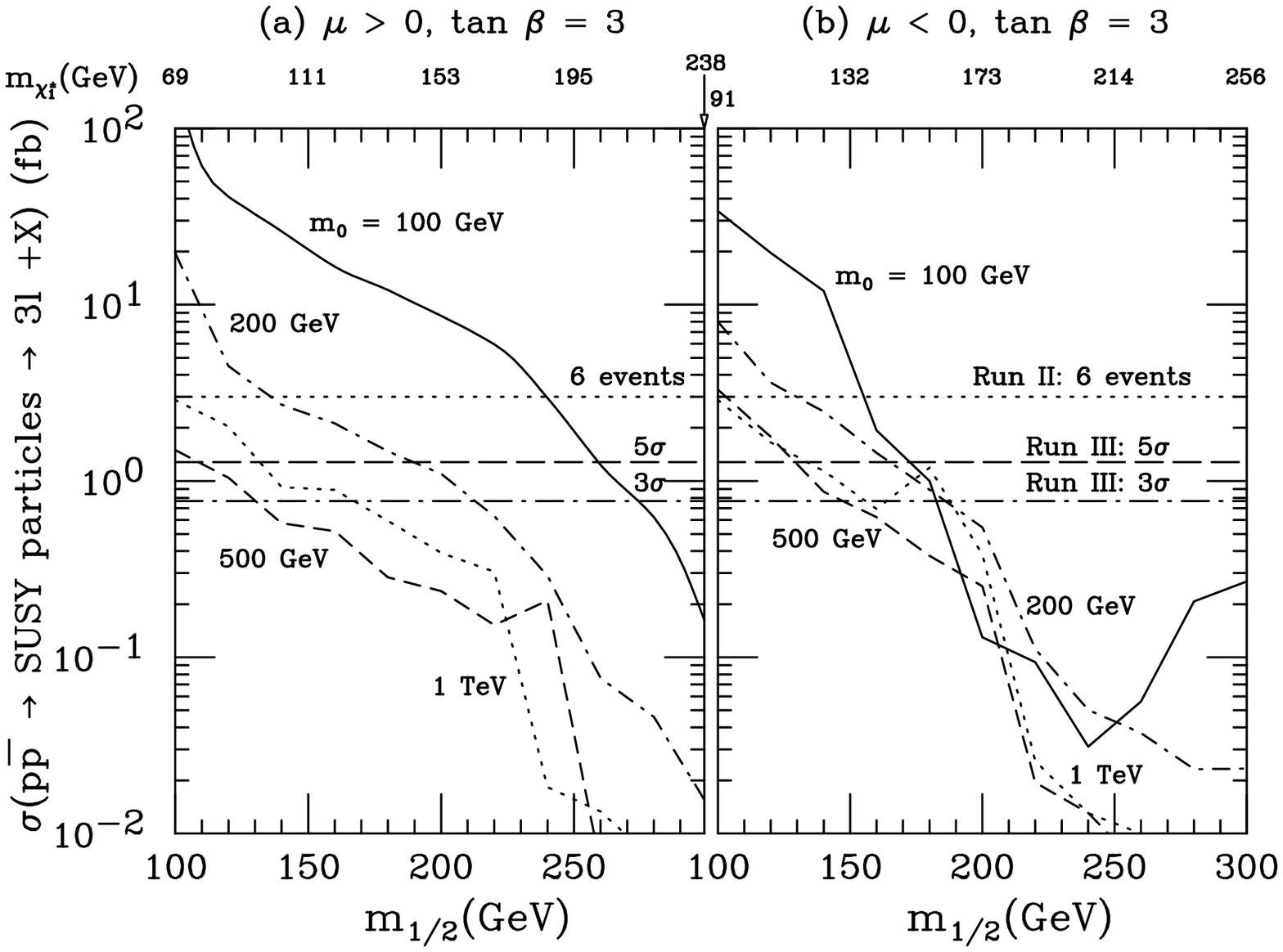}

\bigskip
\caption[]{
Cross section of $p\bar{p} \to {\rm SUSY \; particles} \to 3\ell +X$
with soft acceptance cuts [Eqs. (\ref{eq:Basic}) and (\ref{eq:cuts2})],
versus $m_{1/2}$, at $\sqrt{s} = 2$ TeV,
with $\tan\beta =$ 2, $m_0 =$ 100 GeV (solid), 200 GeV (dot-dash),
500 GeV (dash) and 1000 GeV (dot) for (a) $\mu > 0$ and (b) $\mu < 0$.
Also noted by lines are the cross sections for
(i) 6 signal events with ${\cal L} =$ 2 fb$^{-1}$ (dot), and
(ii) 5 $\sigma$ signal (dash) as well as
3 $\sigma$ signal (dot-dash) for ${\cal L} =$ 30 fb$^{-1}$.
The chargino mass is given on the top horizontal scale with $m_0 = 500$ GeV.
\label{fig:xmhf1}
}\end{figure}
%

\begin{figure}
\centering\leavevmode
\epsfxsize=6in\epsffile{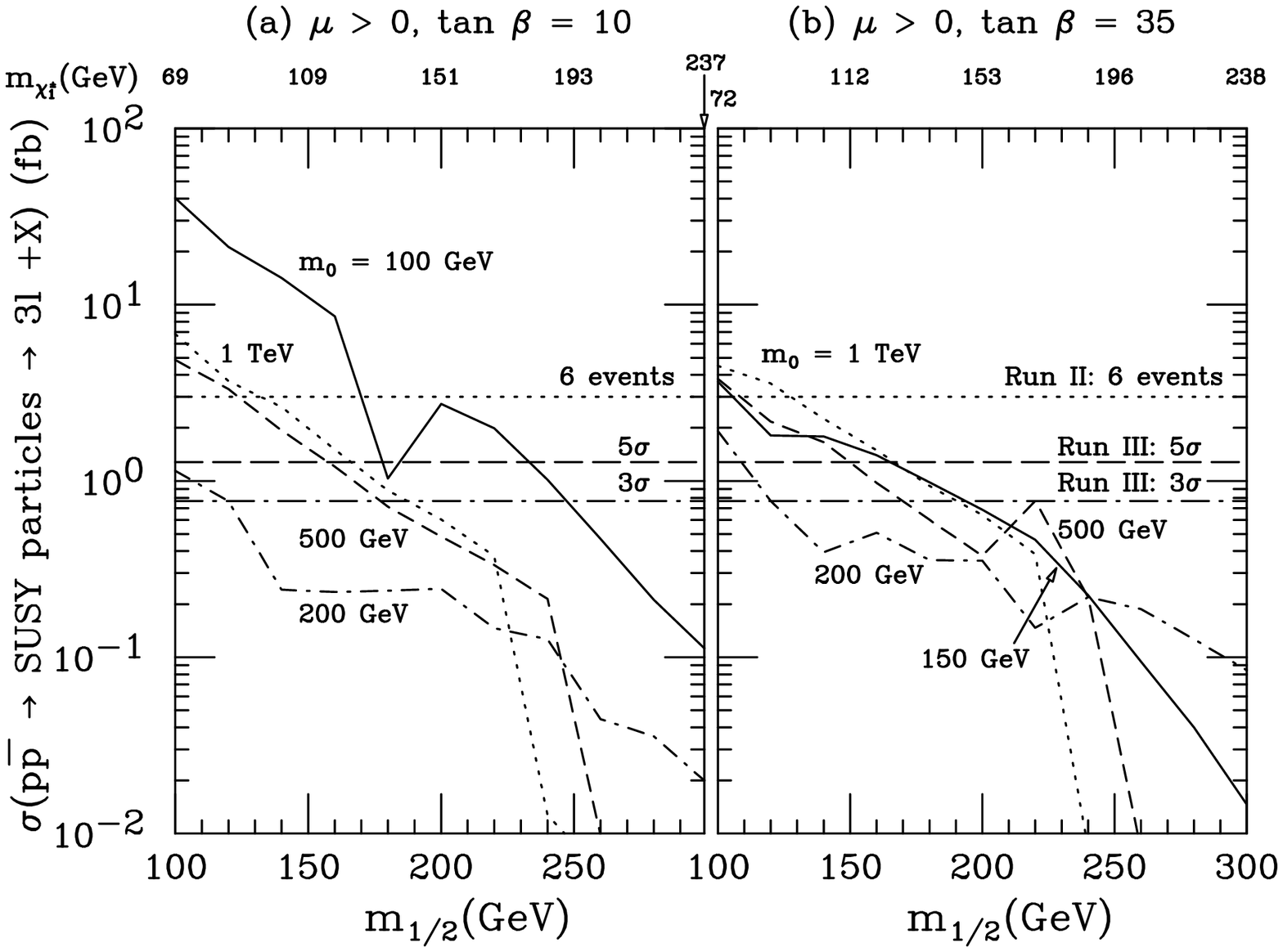}

\bigskip
\caption[]{
Cross section of $p\bar{p} \to {\rm SUSY \; particles} \to 3\ell +X$
with soft acceptance cuts [Eqs. (\ref{eq:Basic}) and (\ref{eq:cuts2})],
versus $m_{1/2}$, at $\sqrt{s} = 2$ TeV,
with $\mu > 0$, $m_0 =$ 100 GeV (solid), 200 GeV (dot-dash), 500 GeV (dash)
and 1000 GeV (dot),
for (a) $\tan\beta =$ 10 and (b) $\tan\beta = 35$,
Also noted by lines are the cross sections for
(i) 6 signal events with ${\cal L} =$ 2 fb$^{-1}$ (dot), and
(ii) 5 $\sigma$ signal (dash) as well as
3 $\sigma$ signal (dot-dash) for ${\cal L} =$ 30 fb$^{-1}$.
The chargino mass is given on the top horizontal scale with $m_0 = 500$~GeV.
\label{fig:xmhf2}
}\end{figure}
%

\begin{figure}
\centering\leavevmode
\epsfxsize=6in\epsffile{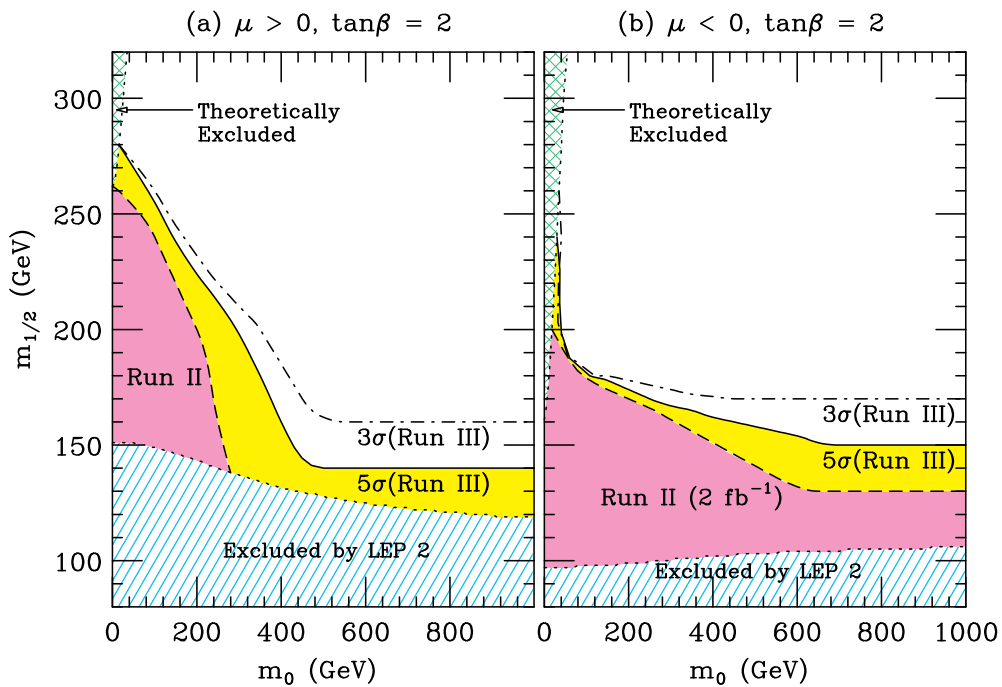}

\bigskip
\caption[]{
Contours for 99\% C.L. observation at Run II and
5$\sigma$ discovery as well as 3$\sigma$ observation at Run III
in the $(m_0,m_{1/2})$ plane,
for $p\bar{p} \to {\rm SUSY \; particles} \to 3\ell +X$
at $\sqrt{s} = 2$ TeV with soft acceptance cuts
[Eqs. (\ref{eq:Basic}) and (\ref{eq:cuts2})],
for $\tan\beta =$ 2, (a) $\mu > 0$ and (b) $\mu < 0$.
All SUSY sources of trileptons are included.
The shaded regions denote the parts of the parameter space excluded by
(i) the theoretical requirements, or (ii) the chargino search at LEP 2.
\label{fig:contour2}
}\end{figure}
%

\begin{figure}
\centering\leavevmode
\epsfxsize=6in\epsffile{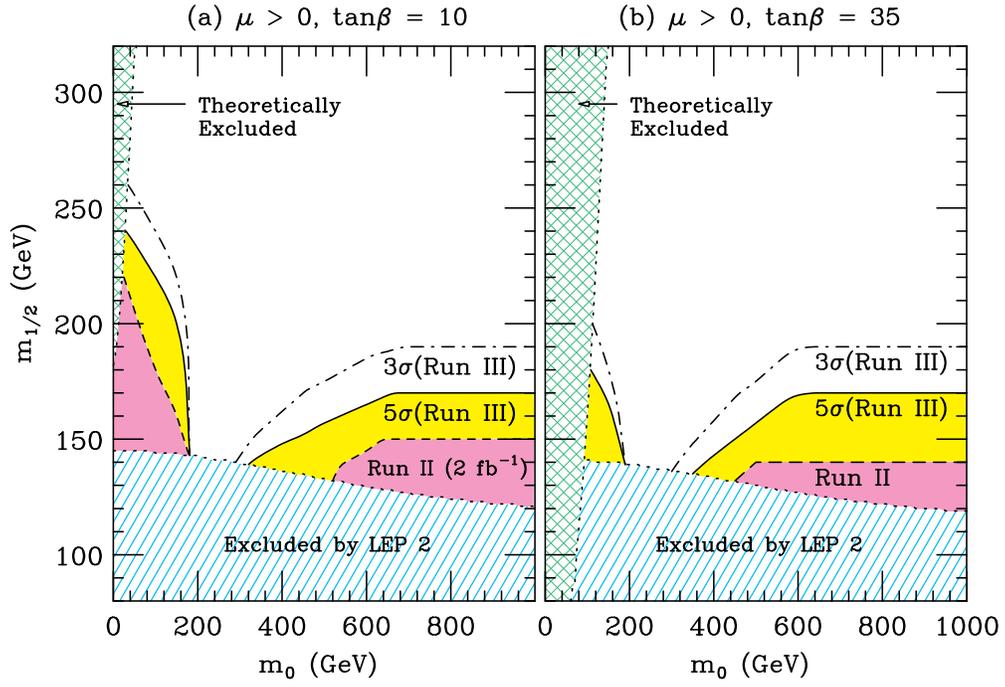}

\bigskip
\caption[]{
The same as Fig. \ref{fig:contour2},
for $\mu > 0$, (a) $\tan\beta =$ 10 and (b) $\tan\beta = 35$.
\label{fig:contour10}
}\end{figure}
%

\begin{figure}
\centering\leavevmode
\epsfxsize=6in\epsffile{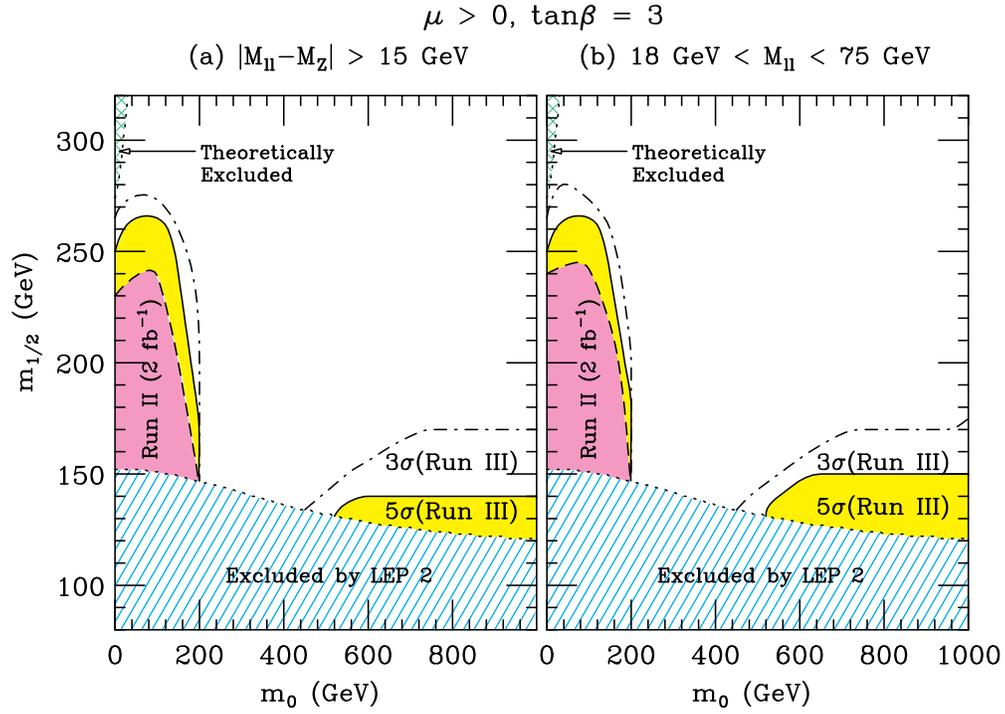}

\bigskip
\caption[]{
The same as Fig. \ref{fig:contour2},
for $\mu > 0$ and $\tan\beta = 3$,
with (a) soft cuts A ($|M_{\ell\bar{\ell}}-M_Z| >$ 15 GeV)
and (b) soft cuts B (18 GeV $\leq M_{\ell\bar{\ell}} \leq$ 75 GeV).
The calculations in this figure are based on ISAJET 7.44
which incorporates the decay matrix elements
for the charginos and the neutralinos.
\label{fig:contour3}
}\end{figure}

\begin{figure}
\centering\leavevmode
\epsfxsize=6in\epsffile{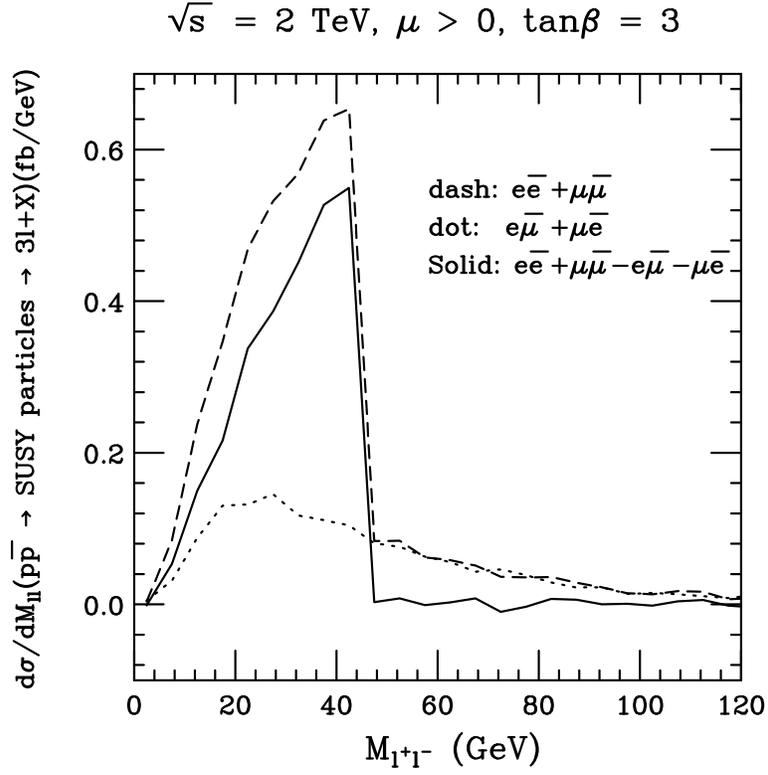}

\bigskip
\caption[]{
The subtracted invariant mass distribution
for the same flavor lepton pairs with opposite signs ($\ell^+\ell^-$)
as defined in Eq. \ref{eq:xmll},
for $p\bar{p} \to {\rm SUSY \; particles} \to 3\ell +X$
at $\sqrt{s} = 2$ TeV,
with $\mu > 0$, $\tan\beta =$ 3, $m_{1/2} = 200$ GeV, and $m_0 =$ 100 GeV.
\label{fig:xmll}
}\end{figure}

\end{document}